\documentclass[]{interact}

\usepackage{epstopdf}
\usepackage[caption=false]{subfig}

\usepackage[natbibapa, nodoi]{apacite}
\usepackage[colorlinks=true,citecolor=black,linkcolor=black]{hyperref}

\theoremstyle{plain}

\theoremstyle{definition}

\theoremstyle{remark}

\newcommand{\citeshortp}[1]{\shortcites{#1}\citep{#1}}
\newcommand{\citeshortt}[1]{\shortcites{#1}\citet{#1}}

\usepackage{hyperref}
\usepackage{cleveref}
\usepackage{listings}
\usepackage{float}
\usepackage{graphicx}

\begin{document}

\title{Confronting Conflicts to Yes: Untangling Wicked Problems with Open Design Systems}

\author{
\name{L.G. (Lukas) Teuber \textsuperscript{a,}*\thanks{* Email: lukas.teuber@boskalis.com.}, A.R.M. (Rogier) Wolfert \textsuperscript{a,b,}*\thanks{* Email: rogier.wolfert@odesys.nl.}}
\affil{
\textsuperscript{a} Decision Science \& Engineering at Boskalis: boskalis.com.
}
\textsuperscript{b} R\&D at ODESYS: odesys.nl.
}

\maketitle

\begin{abstract}
Current project development practices often fail to engage stakeholders early and effectively. Decision support is often non-inclusive, single-sided, and lacking in transparency, while complexity goes beyond human’s comprehension. Additionally, many approaches focus primarily on technical system aspects, neglecting the integration of stakeholders’ individual preferences. This often results in project impasses, leaving stakeholders unable to collaboratively achieve a "yes."
There is a need for a purely associative, a-priori design approach that integrates system realities and stakeholder ideals within a joint socio-technical solution space. The state-of-the-art Preferendus, embedded in the proven Open Design Systems (Odesys) methodology, is a neutral tool for transforming complexity into success. Aiming for synthesis, Odesys’ robust IMAP optimization method generates a single best-fit design solution.  
Here, Odesys is applied for a Dutch wind farm stalemate development, balancing multiple stakeholder preferences, wind farm performances, and project constraints. The success of this approach hinges on stakeholder trust and input. This article introduces a structured stakeholder assessment method using choice-based conjunctive analysis (CBCA), facilitating transparent determination of global and local stakeholder weights and preference functions. 
Modelling ‘disputable’ exogenous factors as endogenous design parameters, the application demonstrates how one can shift toward a collaborative "yes." For this, it is concluded that a zoomed-out solution space would enable the energy transition to be tackled with multiple options rather than a prescribed one. The Odesys approach fosters decision-making that aligns with the social threefold principles of freedom, equality, and fraternity, guiding projects toward genuine democratic outcomes rather than selecting from curated options.
\end{abstract}


\section{Introduction}

\subsection*{Context}

Why do we often build what nobody wants? Because in complex developments, we involve stakeholders too late, or sometimes not even at all \footnote{In this context, note Lucebert's famous line of poetry: 'everything of value is defenceless'}, the decision-making process is not transparent and the creative process is not best-fit for common purpose. In fact, the stakeholder is often only allowed to choose after the fact, as a kind of ‘cosmetic dressing’, from variants conceived by someone else that are not certain to be optimal but allow room for manipulation. And, why are stakeholders digging themselves ever deeper into a ‘technical-scientific swamp’ while failing to get out of these stalemates conjointly? Actually, in this way they keep battling for a presumably biased and sub-optimal alternative. And, why do (political) decision-makers usually hold the cards to their chests without allowing all stakeholders to participate equivalently and objectively? As such, planning and development processes often end in a deception for stakeholders and synthesis will not emerge. Would it be not great to associatively arrive at a best-fit and integrative socio-technical solution, aided by open and neutral computer glass-box modelling? But now, how do we take a co-creative route to a conjoint ideal within reach? 

Until recently, science also offered no solace for this route of open design and decision-making, because decision models either focus only on what people desire or only on what is feasible, and often focus only on the technique of mathematical optimisation rather than integrating it into the art of human-centred design. The state-of-the-art Open Design Systems (read Odesys and see \citeshortt{Wolfert2023}) methodology however does offer an opportunity to openly overcome conflicts of socio-technical interest. This multi-objective design optimisation (MODO) methodology Odesys ideally unites wishful thinking (‘desireabilty’: 'what one wants') and realism (‘capability’: 'what it can do' ). Odesys makes all interests count, resulting in a co-created best-fit solution, where value is more than money or technology alone.  Through an open glass-box modelling in which this integration is maximised for the associated group preference (IMAP), Odesys determines a best-fit synthesis using its solid mathematical design engine the Preferendus, which acts as an extra 'perceptual organ' in the development process. Odesys and its IMAP/Preferendus is a pure form of complex systems integration and design participation to enable an optimal co-creative synthesis unlocking from the beginning rather than sub-optimal compromises after the fact. In other words, from insight outwards and from outlook inwards, Odesys has a way out to a conjoint ideal within reach. 

Odesys works from the design principle of associative \& integrative \textit{diaduction} \footnote{Diaduction (compare words dialogue, dialectic) as opposed to the specific Eeekels\&Roozenburg term 'innoduction' (see \citeshortt{Wolfert2023}) : i.e., it is not the innovation from the design process which is predominant but the transformation of a socio-technical conflict into co-creative success. One can  consider this process as a diaduction of spiritual mind, physical matter, outer observation and inner experience. In other words, from the social participation and technical performance point of view both subject and object are integrated and individual stakeholder preferences are associated. Note: the term diaduction is a completely new term linked to the specifics of a design process, which relates to the concepts of dialogue and dialectics (with a link to the term diacracy, see next section), but which indicates a pure open and dynamic engineering approach aimed at a best-fit value, as a 'contramal' / counterpart to the dialectical induction within research which pursues unique truth. Design enables an integrative way out (ie., duction) of a congrgate of multiple (ie., dia) 'interests' and 'degrees of freedom' (compare terms as induction, deduction and see \citeshortt{Wolfert2023})} and is able to resolve a design freedom paradox. In fact, designing is solving a freedom paradox in which from the individual human freedom \footnote{When we describe creative activity, we speak of freedom of style or freedom of self-expression in a way that indicates an inner conquest of outer constraints. An inner degree of freedom implies something granted or imposed from outside’s degrees of freedom and constraints. Designing is thus an inner-outer spiritual activity, on action, on thinking and feeling that emerges from the human inner self to materialise in the outer world the maximum potential of a the system freedom is found. Here, this maximum can be synthesised when the individual makes less claim to his own self-interest in the interest of the well-being and potential of the whole, see Odesys' social-technical threefolding principles (see \citeshortt{Wolfert2023})} and the artefact's design freedom are maximized for the highest associated \footnote{Associated originates from the term 'association', which is similar terms are covenant, alliance (UK, French) or Bundnis (GER). It indicates co-existence and fraternity in which individual interests are aggregated. A pure association presupposes holding on to each other out of trust and a willingness to 'cross over in a common boat despite not knowing exactly where you are going end up.} group value. Moreover, Odesys computer-aided design-support methodology aligns also with \cite{Wolfert2023} "thinking slow" (or 'deliberative thinking') using simulation and optimization models to explore numerous solutions systematically, reducing cognitive biases, enhancing efficiency, and acknowledging results no longer conceivable for humans. It thus facilitates a shift within complex systems design management from unsubstantiated 'thinking fast' to transparent 'thinking slow' glass-box modelling. This mode of a-priori and joint decision support modelling is one of deliberation (which means literally: to free as a whole). Thereby, both the social and technical cycles are supported to open a best-fit-for common purpose.

These advances culminate in a holistic and open design methodology which integrates the system degrees of freedom ('object') and the individual freedom of the project stakeholder ('subject') seeking for a group maximisation of the associated preferences of the stakeholders involved, within a constrained design space. This generative design freedom maximisation can be seen as an open design aggregation of 'vertical' integration and 'horizontal' association and aligns with Odesys' social-technical threefold design principles of individual design freedom, human equality, and stakeholder fraternity, promoting a purposeful and balanced project compass (see \citeshortt{Wolfert2023}). With Odesys and its U-model, a next step is taken in computer-aided decision science and engineering to optimally confront complex design freedom toward a collaborative yes. It represents a shift in decision-making, where the paradigm of allegedly free choice — which is actually a selection from curated options — shifts into genuinely choosing from a neutral space of 'infinite' freedom, ultimately arriving at a best-fit synthesis for a common purpose within socio-technical reach.

\subsection*{Contribution}
Projects where stakeholders (politicians, residents, environmental organisations, project developers etc.) find it difficult to reach a synthesis are the complex development challenges of onshore windfarms. Here, energy returns and long-term effects for people's well-being and a healthy living environment are at odds. The plan developers responsible are municipalities and other government bodies that present preconceived 'preferred alternatives' at public consultation evenings and try to push these through, without the stakeholders really being heard from their usual NIMBY syndrome. Engineering consultants and other advisors often benefit financially on a stalled process and are willing to generate new alternatives indefinitely, not realising enough that they are missing opportunities for even better best-fits. Above all, they often engage in a purely technical discourse, where a explorative model becomes an opinion in a mathematical form rather than integrating the different opinion into a normative model. Moreover, they never realize that their puzzle as a whole was one piece of a larger puzzle. Odesys and its IMAP based Preferendus can open up this 'puzzle process' and turn this conflict of interest into a co-creative 'yes' both effectively and efficiently. This novel design approach will turn decision-making upside down, steering it toward a truly direct-democratic form by: (1) reversing sub-optimal compromises made after the fact to achieve a single design synthesis in one go; (2) rotating from a vertical, top-down hierarchy to a horizontal, networked association; and (3) shifting from a technical, single-sided view to a holistic, human-centred perspective that integrates idealism with realism." Based on a real-life demonstration project of a wind farm in the Dutch municipality of Oss, the Odesys methodology will further elucidate how to structure a stalemate problem and enable it to be unlocked. The specific contributions of this paper are (1) translating and resolving a stalemate problem into a multi-objective socio-technical threefolding design problem in which uncontrollable variables can be converted to controllable variables to smooth the problem from a purely single-side technical stalemate; (2) a conjoint analysis based preference elicitation for determining individual global and local stakeholder weights as well as a first estimation of their preference functions. Finally, this stalemate design application (DA) can be seen as a complement to the summative Odesys DAs from \citeshortt{Wolfert2023}.

 \subsection*{Structure}
First, in Section 2, the mathematical Odesys threefold formulation will be used to formulate a socio-technical problem as an a-priori design problem. This requires an integrative set of performance, objective and preference functions which will be addressed in Section 3 for the wind farm problem. The design variables and the related design space constraints  are also introduced. Next in Section 3, a new structured stakeholder judgement approach is introduced for preference elicitation, using the a choice-based conjoint analysis (CBCA) method. In Section 4, the results and a conspection will be presented for three cases: (1) a simplified linear 2x2 multi-objective and single interest design problem (MODO) explaining a compromise solution for equal global weights within a plottable solution space; (2) a linear 4x4 multi-objective and single interest design problem (MODO) explaining a synthesis solution for both an equal and uneven distribution of the global weights; (3) a non-linear 4x4 multi-objective and multiple interests design problem (MODO) explaining a synthesis solution having both global and local weights and interests. Section 5 summarises the main conclusion and next steps for further developments. 

 \subsection*{Outreach }
Despite the presented 4x4 multi-objective and multiple interests modelling being a direct reflection of the actual complex design problem, achieving real results requires iterative collaboration with all actual stakeholders. In addition to the serious game results as presented here, this requires the use of the Odesys U-model that facilitates the technical, social and purpose process by going through open design loops \citeshortp{Wolfert2023}. This is beyond the scope of this paper. The focus of this paper is therefore on providing and demonstrating a design loop for several cases, rather than a specific and detailed treatment of participatory decision-making of the complex wind farm project as such. The authors are convinced that the approach presented here makes it possible to renew the current ‘pseudo-democratic’  \footnote{Actually, the word pseudo-democracy is used because nowadays the ‘stakeholders’ are kept happy but not really involved. That is why, as a counter-movement, all kinds of forms such as 'deep- or direct-democracy' are emerging. However, none of them has an operationalisation of decision support modelling based on their ‘well-intentioned’ principles. Therefore, it is time to adopt a socio-technical threefolding design modelling approach that fuses freedom, fraternity and equivalency.} form of decision-making towards a design diacracy for the common well-being from the future \citeshortp{berard2023}. This includes a shift from a Referendum (ex-post) towards a Preferendum (ex-ante) based decision making resulting in healthy solutions for all concerned making use of socio-technical threefolding principles \citeshortp{Wolfert2023}. Finally, this is also a call to the many different engineering consultants and contractors supporting complex projects to broaden their horizons and further improve their current best practices to achieve transparent, effective and efficient decision-making. From this basic open design attitude, the Odesys approach be of added value in complex projects to confronting conflicts to a co-created yes by informed decision-making that eliminates biases.

 \subsection*{Reading note}
For the reader here are two more important general notes (1) this article contains only additional references to those in the Open Design Systems book, see \citeshortt{Wolfert2023}. However, we would like to draw the reader's attention specifically to a number of works from this extensive list which are: Ackoff, Glasl, Roozenburg\& Eekels and Scharmer. This article contains new developments with respect to their work. (2) the three case results are only a subset of the real-life design process and therefore only serve to qualitatively explain the Odesys methodology and the open-ended U-process rather than discussing the quantitative results as such. The design outcomes are the result of a serious game which took place within Boskalis. The actual outcomes are confidential, but the subset as presented here are illustrative.

\section{ Odesys' mathematical threefold formulation}\label{sec:optimisation}
The purpose of this section is to enable stand-alone reading of this paper. To this end, the Odesys is summarised here (for more info see \citeshortt{Wolfert2023}; \citeshortt{Teuber2024}; \citeshortt{VanHeukelum2024}). The core of Odesys' methodology is the following mathematical statement of a multi-objective design optimisation (MODO) problem, which integrates subject desirability and object capability and generates a feasible solution with the maximised aggregated group preference: i.e. the IMAP method, and reads as 

\begin{equation}
    \label{eq:general_MS}
    \begin{gathered}
    \mathop{Maximise}_{\mathbf{x}} \textrm{ } U = A \left[  
    P_{k,i} 
    \left(O_i
    \left(F_1(\mathbf{x}, \mathbf{y}),F_2(\mathbf{x}, \mathbf{y}),...,F_J(\mathbf{x}, \mathbf{y}) \right)
    \right),w'_{k,i} \right] \textrm{ for } \\ 
    k=1,2,...,K \\ 
    i=1,2,...,I
    \end{gathered}
\end{equation}

\noindent
constrained by: 
\begin{equation}
    \label{eq:ineq_cons}
    \begin{gathered}
    g_{p}(O_{i}(F_{1,2,...,J}(\mathbf{x}, \mathbf{y})),F_{1,2,...,J}(\mathbf{x}, \mathbf{y})) \le 0
    \textrm{ for } p=1,2,...,P
    \end{gathered}
\end{equation}

\noindent
and where:

\begin{itemize}
    \item $U$: Utility function that needs to be maximized for a best-fit design configuration, where the Preferendus Genetic Algorithm is used for \citeshortp{Wolfert2023} or \citeshortt{VanHeukelum2024}.
    
    \item $A$: An algorithm that determines the aggregated preference score as part of the IMAP optimisation operation (see \citeshortt{Teuber2024} or the  \citeshortt{boskalispython2024}).
    
    \item $P_{k,i}(O_i(F_{1,2,...,J}(\mathbf{x}, \mathbf{y})))$: Preference functions that describe the preference stakeholder $k$ has towards objective function $i$, which are functions of different design performance functions $j$ and dependent on controllable design and/or non-controllable physical variables ( $i \le k$ and $K$ is maximum number of stakeholders).
    
    \item $O_i(F_{1,2,...,J}(\mathbf{x}, \mathbf{y}))$: Objective functions that describes the objective $i$.
    
    \item $F_{1,2,...,J}(\mathbf{x}, \mathbf{y})$: Design performance functions $j$ that describe the physical object behavior.
    
    \item $\mathbf{x}$: A vector containing the (controllable endogenous) design variables $x_1,x_2,...,x_N$. These variables are bounded such that $lb_n \le x_n \le ub_n$, where $lb_n$ is the lower bound, $ub_n$ is the upper bound, and $n=1,2,...,N$.
    
    \item $\mathbf{y}$: A vector containing the (uncontrollable exogenous) physical variables $y_1,y_2,...,y_M$.
    
    \item $w'_{k,i}$: Weights for each of the preference functions. The global weights for the relative importance of stakeholders is defined as $w_{k}$. The local weight stakeholder $k$ gives to objective $i$ is defined as $w_{k,i}$. The following formula holds: $w'_{k,i}=w_{k}\cdot w_{k,i}$, given that $\sum w'_{k,i}=\sum w_{k,i}=\sum w_k=1$. Note that: (1) in case of equivalent stakeholder decision-making \( w_k = 1/K \), (2) these global/local weights within our MODO problem are similar to main and sub-criteria in an MCDA evaluation, see \citeshortt{Wolfert2023} and (3) the maximum number of preference functions equals $K*I$.

    \item $g_{p}(O_{i}(F_{1,2,...,J}(\mathbf{x}, \mathbf{y})),F_{1,2,...,J}(\mathbf{x}, \mathbf{y}))$: Inequality constraint functions, which can be either objective function and/or design performance function constraints. Note: should there be any equality constraints in the stalemate problem, these will be rewritten as an inequality constraint $g_p$.

\end{itemize}

\begin{figure}[H]
    \centering
    \includegraphics[width=0.6\linewidth]{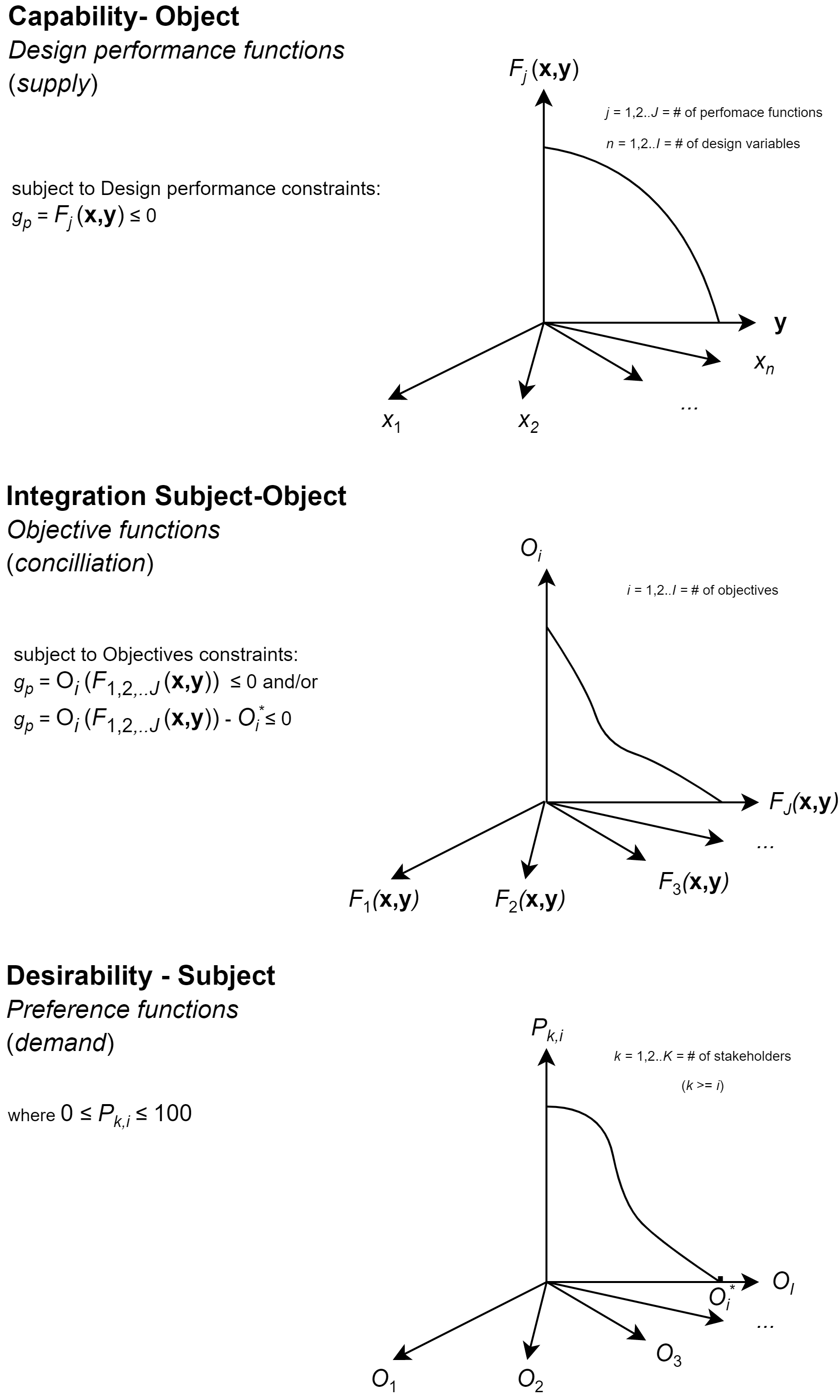}
    \caption{Conceptual threefold framework of the Odesys mathematical statement, where subject desirability and object capability are integrated, see \citet{Wolfert2023}. Note that: (1) the shapes of the curves are arbitrary; (2) the different functions are linked (an ordering principle) and that maximisation is not yet part of this threefold. }
    \label{fig:threefold}
\end{figure}

To better understand and further detail this specific social-technical systems integration, the different functions as part of the mathematical formulation are conceptualised in the Odesys threefold modelling framework, as shown in \autoref{fig:threefold}.

\section{Wind farm development - a stalemate case}

First we will describe here the socio-technical wind farm design stalemate problem qualitatively, as a basis to translate into a mathematical MODO formulation with corresponding quantitative IMAP design outcomes.

\vspace*{10pt}

\noindent
\textbf{\textit{ Technical context}} This wind farm design project is to generate renewable energy for the city Oss. To meet its energy goals, the municipality wants to develop a wind farm with a number of turbines, which should ideally be as close to the center as possible to minimize energy losses. It is estimated that no more than 12 turbines are needed. Energy generation depends not only on the number and location of the turbines but also on the turbine height and the average wind speed required, among other factors. These four design variables can still be controlled. The current search c.q. preferred area is located on a (radius) line from the center of Oss to the west and extends to 10km inside the municipal border. The wind farm should be located at least 2km outside the center of Oss.  The height of the turbines will not exceed 150m (hub height). Typical uncontrollable variables are: rotation diameter, type of power converter, specific local environmental data etc.
The different design performance functions depend on the aforementioned design variables. Below are the relations for three performance functions between power generation; noise hindrance and bird mortality on the one hand and the four design variables on the other.

\noindent
\textbf{\textit{ Social context}}
This wind farm design as part of sustainable polder development, demonstrates a multi-stakeholder design optimisation
approach for the decision on the number and type of turbines at a certain location. The creative conflicts of interests are : (a) energy - profit for the energy service provider; (b) health- noise hindrance for the local residents; (c) environment - bird mortality for the ecologists; (d) health - particle pollution for regulator RIVM; amongst others such as 
health \& environment - magnetic cable radiation, horizon-pollution, and shadow cast. The energy service provider's objective is to maximise the energy power and it's profit within a positive business-case. The objectives of all other stakeholders should be minimised. From its contribution to the national energy transition, the municipality of Oss has a minimum supply requirement of renewable energy ('RES-bijdrage'). 

\noindent
\textbf{\textit{ Stalemate Context}}:
Here are two actual quotes which clearly descibes the stalemate (see 'Wind farm rage Oss' \citeshortt{blckbxtv2024})

\noindent
Municipality : "The current preferred alternative is the "best plan" according to the Municipality, but it remains a "compromise": it will never succeed in satisfying everyone. It remains a balancing of interests. We will all have to contribute to the energy objective, and as we have now made the calculations we will need 9?"

\noindent
Other Stakeholders : "The alternative does not fit in any way: not within the framework, not within the energy objective. It is wishful thinking, without realism. The current plan has only losers. According to other Stakeholders there will be actions. We are 6-0 behind, and we do not have the resources like the Municipality, developers or a collective of farmers. We will take action unless the Municipality really comes to talk to us now. We have nothing against windmills in principle.  

\vspace*{10pt}

Based on this contextual description, we will now describe the integrative MODO problem by running through the Odesys threefold mathematical statement framework (see Section 2), resulting in design performance-, objective-, and preference functions (see Section 3), including the design degrees of freedom (design variables and stakeholder's preference and importance).

\subsection{Design performance functions}
In this section, we will introduce all the relevant design performance functions for a wind farm. These functions describe the wind farm behaviour as a function of design variables or other physical non-controllable variables and they will be integrated into the objective functions in Section 3.2. Here, we have used \cite{engineeringtoolbox2009wind} and \citeshortt{Kalmikov2017} generically and refer to additional specific references per performance function. The following design variables 
 and their design space bounds are considered. 

\begin{enumerate}
    \item $F_1=x_1 \text{\ } (2\le x_1 \le10)$: distance to city centre
    \item $F_2=x_2 \text{\ } (0\le x_2 \le12)$: number of turbines
    \item $F_3=x_3 \text{\ } (50\le x_3 \le150)$: turbine (hub) height
    \item $F_4=x_4 \text{\ } (3\le x_4 \le15)$: required wind speed
\end{enumerate}

\subsubsection{$F_5 = P_{dt}$ = power generated per day per turbine}

To calculate the power generated per day per turbine, we start with the blade diameter calculation. The diameter \(D\) is given by:
\begin{equation}
D = x_3 \times 1.3
\end{equation}
where \(x_3\) is the turbine (hub) height and the constant (1.3) is a scaling factor. This factor is derived from empirical data and/or industry standards to account for certain physical characteristics or performance metrics of the turbine.

Next, the swept area \(A\) of the wind turbine blades is calculated using the formula:
\begin{equation}
A = \pi \left( \frac{D}{2} \right)^2
\end{equation}

The power \(P\) generated per day is then calculated using the following equation:
\begin{equation}
P = 0.5 \times A \times \rho \times e \times (x_4)^3
\end{equation}
where \(A\) is the swept area, \(\rho\) is the air density (1.225 kg/m³), \(e\) is the efficiency (0.3), and \(x_4\) is the average rated wind speed. The power \(P\) is in Watts.

Finally, to convert the power from Watts to kilowatt-hours (kWh) per day, we use:
\begin{equation}
P_{dt} = P \times \frac{24}{1000}
\end{equation}
Note one could introduce an inequality constraint here, e.g., \( g_1 \geq 50000\) kWh/day (minimum energy supply requirement).

\subsubsection{$F_6 = SP$ = sound power per turbine}

First of all we have to introduce two adjustments:
1. Wind Speed Adjustment:
\begin{equation}
\text{wind\_adj} = x_4 - \text{wind\_base\_speed} * 0.2
\end{equation}
where \(\text{wind\_base\_speed} = 3\) m/s. This adjustment accounts for an increase in noise by 0.2 dB per m/s over the base wind speed.
2. Height Adjustment:
\begin{equation}
\text{height\_adj} = x_3 - \text{height\_base} *0.1
\end{equation}
where \(\text{height\_base} = 50\) meters. This adjustment accounts for an increase in noise by 0.1 dB per meter over the base height.
Then, the general inverse square law for sound pressure level describes the decrease in intensity of a sound wave as it propagates away from a point source in a free field. From this a general formula for calculating the sound pressure level (SPL) at a distance is calculated by:
\begin{equation}
SPL_{dist} = SPL_{source} - 20 \cdot \log_{10}\left(\frac{dist}{ref\_dist}\right)
\end{equation}
where:
\begin{itemize}
    \item $SPL_{dist}$ is the sound pressure level at the observation point,
    \item $SPL_{source}$ is the sound pressure level at a reference distance from the source,
    \item $dist$ is the distance from the source to the observation point,
    \item $ref\_dist$ is a reference distance from the source, often taken as 1 meter.
\end{itemize}
and adapted for our specific problem:
\begin{equation}
L_{d,\text{single}} = L_W + \text{wind\_adj} + \text{height\_adj} - 20 \log_{10}\left(\frac{x_1 * 1000}{1}\right)
\end{equation}
Here \(L_W = 104\) dB is the example sound power level of a turbine. This formula calculates the noise level at a distance considering the adjustments for wind speed and height. Now we have to convert the single turbine noise level to power using a relationship between sound pressure level and sound power that is expressed as:
\begin{equation}
Power = 10^{\frac{SPL - SPL_{ref}}{10}}
\end{equation}
where:
\begin{itemize}
    \item $Power$ is the sound power ratio relative to a reference power,
    \item $SPL$ is the sound pressure level in decibels,
    \item $SPL_{ref}$ is the reference sound pressure level, typically 0 dB at the threshold of human hearing (20 micro-pascals).
\end{itemize}
and adapted for our specific problem:
\begin{equation}
SP = 10^{\frac{L_{d,\text{single}}}{10}}
\end{equation}
This converts the single turbine noise level from decibels to power. Note one could introduce an inequality constraint here, e.g., \( g_2 \leq 42 \) dB (on the facade).

\subsubsection{$F_7 = M$ = bird mortality}

For the derivation of the design performance function related to bird mortality we refer to \citeshortt{Nilsson2023}. For the blade diameter $D$, we use Equation (3), which is a function of \(x_3\). Moreover, the swept area of the wind turbine blades is calculated by
\begin{equation}
A = \pi \left( \frac{D}{2} \right)^2
\end{equation}
Then, the bird mortality (\( M \)) is estimated as:
\begin{equation}
M = \left( \frac{A}{30000} \right) * C * \begin{cases} 
800 & \text{if } 8 < x_1 < 9 \\
200 & \text{otherwise}
\end{cases}
\end{equation}
where:
\begin{itemize}
  \item A rotor diameter (\( D \)),
  \item A collision probability factor (\( C \)) of 0.005 (or 0.5\%),
  \item A maximum swept area of 30000.
\end{itemize}
Note that there is a bird breeding area located somewhere between 8 and 9 kilometres from the city centre

\subsubsection{$F_8$ = E = erosion rate}
For the derivation of the design performance function related to the erosion rate we refer to \citeshortt{Zhang2023} and \citeshortt{Rimereit2021}. For the blade diameter $D$ and the swept area $A$, we use Equations (3) and (13), which are both a function of \(x_3\). Then the erosion rate is estimated by: 

\begin{equation}
    E = w  * f * r * \left( \frac{A}{30000} \right)
\end{equation}
where:

\begin{itemize}
  \item \(A\): swept area (with \(A < 30000\)),
  \item \(w\): wind impact equal to \((x_4)^3\),
  \item \(r\): rain impact equal to 1 (negligible),
  \item \(f\): pollution probability factor of 0.001 (or 0.1\%).
\end{itemize}
So, this integrates the hub height, wind speed, and number of turbines to calculate the erosion rate on turbine blades due to particle pollution, expressed in terms that are specific to the definitions provided.

\subsection{Objective functions}
Within the stalemate wind farm design problem, there are four conflicting objectives of interest. These functions will now be described as a function of the design variables and the performance functions from Section 3.1.

\subsubsection{$O_1$ = Energy Profit}
The energy efficiency depends on the so-called resistance factor, which is determined as follows: 
\begin{equation}
R = \frac{x_3 + x_4}{12}
\end{equation}
where \(x_3\) is the turbine height and \(x_4\) is the average rated wind speed.
Using the equations for \(R\) and the design performance function \(F_5\), the final objective function which represents the energy profit reads as:
\begin{equation}
O_1 = O_P = (P_{dt} \cdot \text{EP} \cdot x_2) - (R \cdot x_1)
\end{equation}
where:
\begin{itemize}
  \item \text{EP} is the price of electricity per kWh (0.01),
  \item \(P_{dt}\) is the wind turbine profit per day,
  \item \(x_2\) is the number of turbines,
  \item \text{resistance} is as calculated above,
  \item \(x_1\) is the distance to the city centre.
\end{itemize}

\subsubsection{$O_2$ = Noise Disturbance}
Making use of the design performance function \(F_6\), and converting the total noise power 
back to dB, the final objective function which represents the noise disturbance (also sometimes called noise pollution) reads as:
\begin{equation}
O_2 = O_N = 10 \cdot \log_{10}(x_2 \cdot SP)
\end{equation}

\subsubsection{$O_3$ = Bird Mortality}
Making use of the design performance function \(F_7\), the final objective function which represents the bird mortality reads as:
\begin{equation}
O_3 = O_M = M \cdot x_2
\end{equation}
where:
\begin{itemize}
  \item \(M\) is the bird mortality per day,
  \item \(x_2\) is the number of turbines.
\end{itemize}

\subsubsection{$O_4$ = Particle Pollution}
Making use of the design performance function \(F_8\), the final objective function which represents the particle pollution reads as
\begin{equation}
    O_4 = O_{PP} = E * {x_2}
\end{equation}
where:
\begin{itemize}
  \item E is the erosion rate.
  \item $x_2$ is the number of turbines.
\end{itemize}

\subsection{Preference functions \& weights distributions}

\vspace*{10pt}

To make the open design systems approach executable, the last step is to determine the preference functions and weights of the different stakeholder interests. This is a dynamic open-ended approach in reality in which the social and purpose cycles are run through in iterative open loops using the socio-technical  Odesys U-model. The conjoint preference elicitation approach, also called structured stakeholder judgement (compare structured expert judgement for engineering) is an important part of this process. Here, stakeholders define and adjust their individual preference curves and (global/local) weight distributions in relation to their actual ‘idealised design' purpose 'in the now'. In this article we can only show some main steps (and corresponding case results in the next Section 4). To this end, the three following steps within the integrative open-ended design system approach are proposed here: i.e., 
\begin{enumerate}
    \item a \textit{linear 2x2 single-interest} case in which only two stakeholders: the energy provider \(S_1\) and the local residents \(S_2\), each represent a single objective (\(O_1\) and \(O_2\) respectively) with equal importance, making the global weights both $w_k = 0.50$ and the local weights both $w_{k,i}=1.0$. So, each stakeholder has a 100\% single interest in one objective and participates equally. This can build confidence for the working of the system while a compromise solution can (still) be visualised the solution space (2x2 'plottable' space). This results in the following conjoint weight table (note: linear preference curve are introduced in the next section):

\begin{table}[H]
\centering
\caption{Step 1 - Weights for each of the preference functions, according to $w'_{k,i} = w_k \cdot w_{k,i}$. }
\label{tab:weights}
\begin{tabular}{cccccc}
\toprule
 Stakeholder $k$ & $w'_{O1}$ & $w'_{O2}$ & $w_k$ \\
\midrule
S1 & 0.5 & 0.0 &  0.5 \\
S2 & 0.0 & 0.5 & 0.5 \\
\midrule
Total & 0.5 & 0.5 & 1.0 \\
\bottomrule
\end{tabular}
\end{table}    
    ;
    
    \item a \textit{linear 4x4 single-interest} case in which two stakeholders \(S_1\), \(S_2\) (the same as in the previous step (1)) are joined by two other stakeholders, the ecologist \(S_3\) and the RIVM health organisation \(S_4\), reflecting the objective \(O_{1..4}\) respectively. We assume the following two group decision making cases    (a) the energy provider $w_{S1} = 0.50$ “against” a coalition of the other three stakeholders \(S_{2..4}\) who have weights of $w_{S2} = 0.20; w_{S3} = 0.15; w_{S4} = 0.15$ respectively and (b) all stakeholders do have equal global weights of $w_k = 0.25$. In both case the each stakeholder has a 100\% single interest in one objective, which means that all local weights are $w_{k,i}=1.0$. This results in the following conjoint weight tables (note: we model their preference curves again linearly, see next section):

\begin{table}[H]
\centering
\caption{Step2a - Uneven global weights for each of the preference functions, according to $w'_{k,i} = w_k \cdot w_{k,i}$. }
\label{tab:weights}
\begin{tabular}{cccccc}
\toprule
 Stakeholder $k$ & $w'_{O1}$ & $w'_{O2}$ & $w'_{O3}$ & $w'_{O4}$ & $w_k$ \\
\midrule
S1 & 0.50 & 0.0 & 0.0 & 0.0 &  0.50 \\
S2 & 0.0 & 0.20 & 0.0 & 0.0 & 0.20 \\
S3 & 0.0 & 0.0 & 0.15 & 0.0 & 0.15 \\
S4 & 0.0 & 0.0 & 0.0 & 0.15 & 0.15 \\
\midrule
Total & 0.50 & 0.20 & 0.15 & 0.15 & 1.00 \\
\bottomrule
\end{tabular}
\end{table}   

\begin{table}[H]
\centering
\caption{Step2b - Equal global weights for each of the preference functions, according to $w'_{k,i} = w_k \cdot w_{k,i}$. }
\label{tab:weights}
\begin{tabular}{cccccc}
\toprule
 Stakeholder $k$ & $w'_{O1}$ & $w'_{O2}$ & $w'_{O3}$ & $w'_{O4}$ & $w_k$ \\
\midrule
S1 & 0.25 & 0.0 & 0.0 & 0.0 &  0.25 \\
S2 & 0.0 & 0.25 & 0.0 & 0.0 & 0.25 \\
S3 & 0.0 & 0.0 & 0.25 & 0.0 & 0.25 \\
S4 & 0.0 & 0.0 & 0.0 & 0.25 & 0.25 \\
\midrule
Total & 0.25 & 0.25 & 0.25 & 0.25 & 1.00 \\
\bottomrule
\end{tabular}
\end{table}       
    ;
    
    \item a \textit{non-linear 4x4 multiple-interests} case in which the four stakeholders \(S_{1..4}\) (same as in the previous step (2)) are again the represented in the group decision making problem. Here again all there global weights are equal and thereby $w_k = 0.25$, but they can all have individual interest and preferences for multiple objectives \(O_{1..4}\). In other words, each stakeholder can have multiple interests in different objectives, which means that the local weights are are no longer equal to 1.0 (the sum of the local weights per stakeholder should equal 1.0 instead). This results in the following conjoint weight tables (note: we model their preference curves again non-linearly, see next section):

\begin{table}[H]
\centering
\caption{Step 3 - Local Weights for each of the preference functions per stakeholder}
\label{tab:weights}
\begin{tabular}{cccccc}
\toprule
 Stakeholder $k$ & $w_{O1}$ & $w_{O2}$ & $w_{O3}$ & $w_{O4}$  \\
\midrule
S1 & 1.00 & 0.00 & 0.00 & 0.00  \\
S2 & 0.30 & 0.40 & 0.00 & 0.30  \\
S3 & 0.40 & 0.00 & 0.40 & 0.20 \\
S4 & 0.30 & 0.10 & 0.10 & 0.50  \\
\bottomrule
\end{tabular}
\end{table}

\begin{table}[H]
\centering
\caption{Step 3 - Conjoint weights distribution per preference function, according to $w'_{k,i} = w_k \cdot w_{k,i}$. Note that stakeholders may have different individual interests in all four objectives (max. 16), while the overall weights are equally distributed.}
\label{tab:weights}
\begin{tabular}{cccccc}
\toprule
 Stakeholder $k$ & $w'_{O1}$ & $w'_{O2}$ & $w'_{O3}$ & $w'_{O4}$ & $w_k$ \\
\midrule
S1 & 0.250 & 0.000 & 0.000 & 0.000 & 0.25 \\
S2 & 0.075 & 0.100 & 0.000 & 0.075 & 0.25 \\
S3 & 0.100 & 0.000 & 0.100 & 0.050 & 0.25 \\
S4 & 0.075 & 0.025 & 0.025 & 0.125 & 0.25 \\
\midrule
Total & 0.500 & 0.125 & 0.125 & 0.250 & 1.00 \\
\bottomrule
\end{tabular}
\end{table}

\end{enumerate}

For the preferences elicitation in step (3) above, an conjoint and deliberative analysis within a Boskalis confidential serious game (\citeshortt{VanderStee2024}) was used to elicit preference functions and weight estimations by means of standard choice based conjoint analysis (CBCA) software \citeshortt{Sawtooth2007}.  CBCA is a well-known and proven quantitative concept (see e.g. \citeshortt{Raghavarao2010} here applied for estimating both preference functions and weights distributions. This first quantitative step in preference elicitation gives the individual stakeholder insight into the importance of his interest in the different objectives and also provides an initial shape of the associated preference functions. An example for this CBCA apllication for this case are found in the Odesys Toolbox \citeshortt{deepnoteodesys2024} or with more background details in \citeshortt{VanderStee2024}. This first quantitative preference elicitation estimate is followed by a qualitative step in which the stakeholders can freely adjust the CBCA estimates so that he recognizes his idealized design as good as possible in the weights and preferences. Combining the CBCA with with a freely adapted construct of preference functions and weights (for more info for the second part, see \citeshortt{Mol2024}), both quantitative and qualitative approaches are used to ensure an optimal structured stakeholder judgment process as a reflection of the open design.

\section{Results \& conspection}
In this section, we aim to solve the three cases described above as part of the dynamic and open-ended MODO problem. For this, we use the Preferendus tooling in which a best-fit for common purpose design point is generated within the IMAP method using an inter-generational Genetic Algorithm (GA). In other words, the Preferendus finds a point in the solution space what given the specific case parameters and technical reality with the maximum integrative and aggregated preference value for the stakeholder involved. In this way the Preferendus takes us to a common ideal within reach.
In reality, the Preferendus is a dynamic tool that transparently and objectively supports the iterative design and decision-making process consisting of open technical-, social- and purpose loops. We have shown only three 'design snapshots' here as a result. These three were not chosen for the specific outcomes and results but to demonstrate the possibility of breaking open a stalled design, while balancing desirability and capability. The explanation of the results is therefore brief and the reader is further urged to use dynamic toolbox modelling alongside these three 'extreme' cases to dynamically ‘play’ with the model so that it starts ‘talking back’, see \citeshortt{deepnoteodesys2024}.

A specific note is made regarding the wind speed design variable . Normally, this is not modelled as an endogenous variable \textit{x}, but as an exogenous one \textit{y}. Since the Odesys modelling aims to avoid as much technical debate and analysis as possible to generate a socio-technical design solution instead, we choose to retrospectively verify that this found wind speed occurs on average in practice (i.e., verification against past wind profiles). This design approach, including the wind speed reversal, avoids a possible technical stalemate and this can be adjusted at a later design stage. Moreover, it gives therefore an incentive to open stalled technical analyses and conflicts to explore if they might leave design freedom space unexploited.

Two final notes are made regarding the values of the various parameters and the IMAP outcomes: (1) Despite the fact that all socio-technical functions qualitatively reflect reality as well as possible, the values used still need to be verified in practice. This may still influence the current results. (2) Because of the non-linearity of the problem, \textit{the} global IMAP design point cannot be found. It will always be \textit{a} best-fit, and there may be more that are close to such a best-fit. Therefore, we recommend performing an a-posterior MCDA on these possible different best-fit Preferendus IMAP outcomes (note you have to use at least 3 outcomes within this MCDA, see \citeshortt{Wolfert2023}).

\subsection{Case 1: linear 2x2 single-interest}
In these case 1 results, see the quantitative charts and data in the next subsection, we can notice the following high over:
\begin{itemize}
    \item The first two plots of the solution space and design outcomes generated by the Preferendus show the so-called SODO (single-objective design optimisation) ‘corner-points’ for a linear 2x2 problem, best for \(O_1\) or \(O_2\) and their respective stakeholder. 
    \item The third solution space plot shows the so-called compromise solution, which is a third and different MODO 'corner-point', reflecting a maximum number of turbines at a maximum distance from the city centre. Note: with an a-posterior MCDA it can easily be shown that a MODO design point has a higher group value than the SODO points. In other words, relinquishing the pure self-interest, ultimately enables an optimal design solution that benefits the whole group.  
\end{itemize}

\subsubsection{100\% S1}\label{2x2_100_energy}

\begin{table}[H]
    \centering
    \caption{A best-fit design point}
    \begin{tabular}{cc}
    \toprule
    Variable     &  IMAP outcome\\
    \midrule
    $x_1$     & 2 km \\
    $x_2$ &  12 \# \\
    \bottomrule
    \end{tabular}
    
    \label{tab:2x2 energy results}
\end{table}

\begin{figure}[H]
    \centering
    \begin{minipage}[b]{0.32\linewidth}
        \includegraphics[width=\linewidth]{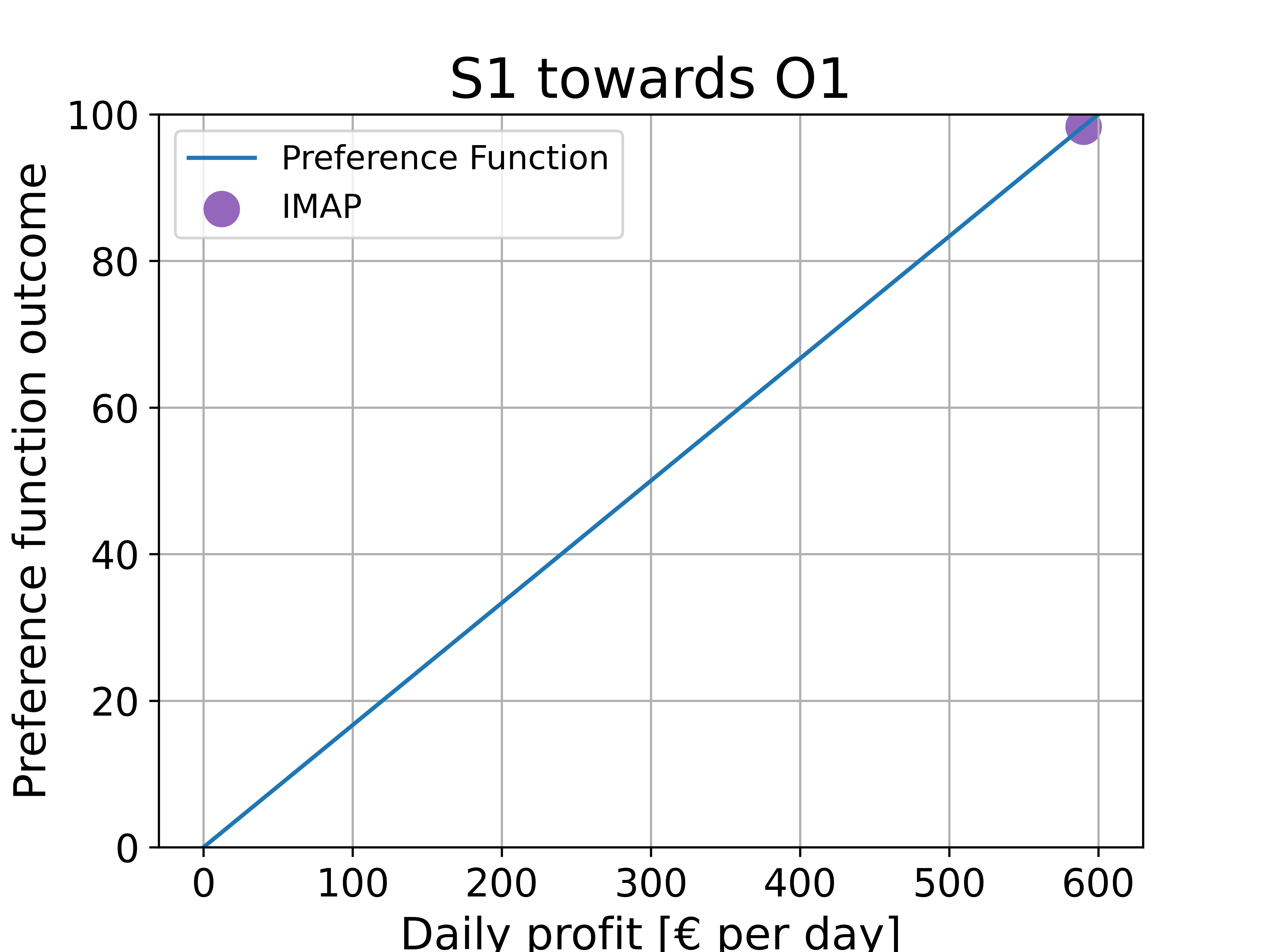}
    \end{minipage}
    \hfill
    \begin{minipage}[b]{0.32\linewidth}
        \includegraphics[width=\linewidth]{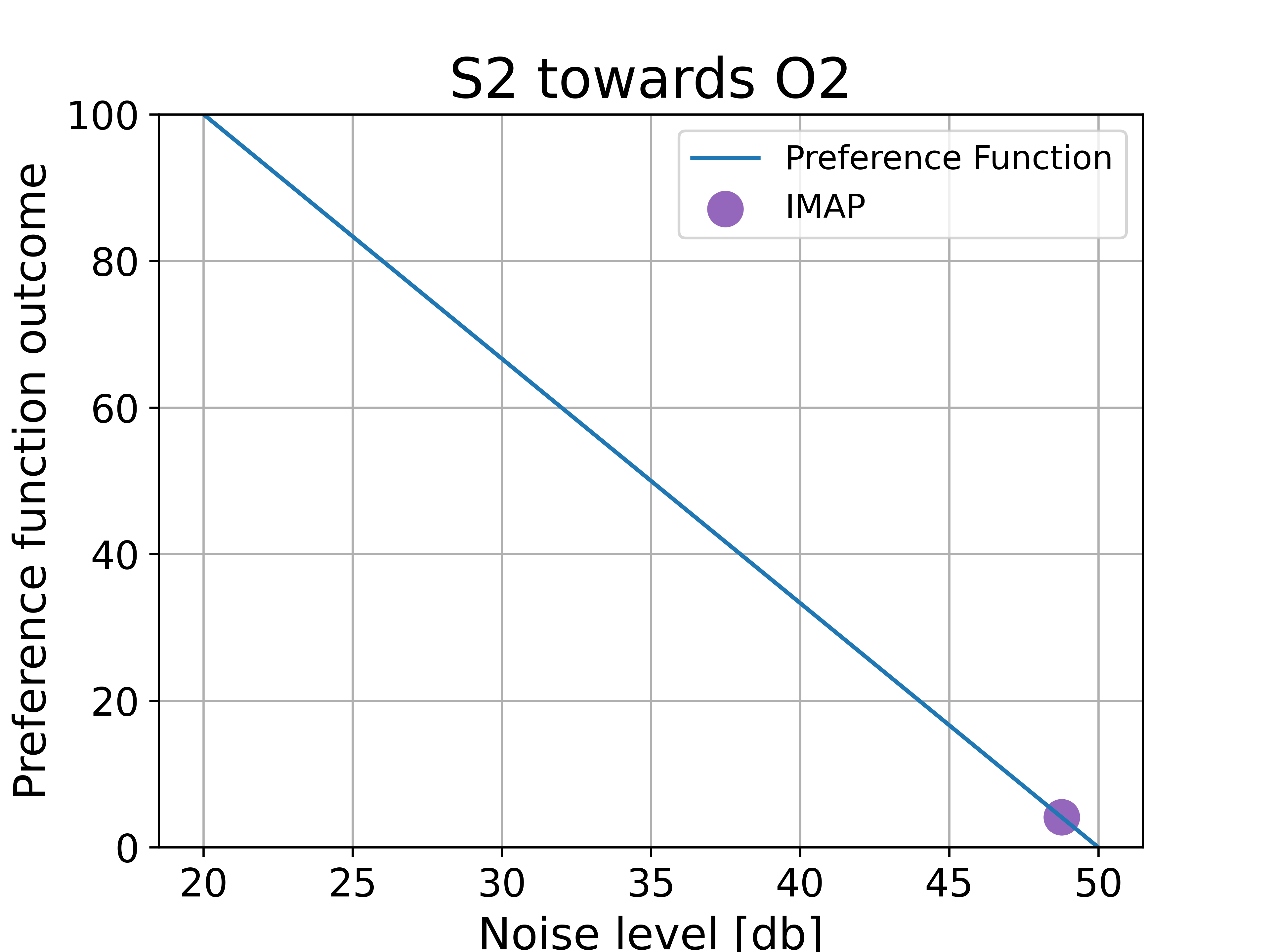}
    \end{minipage}
    \hfill
    \begin{minipage}[b]{0.32\linewidth}
        \includegraphics[width=\linewidth]{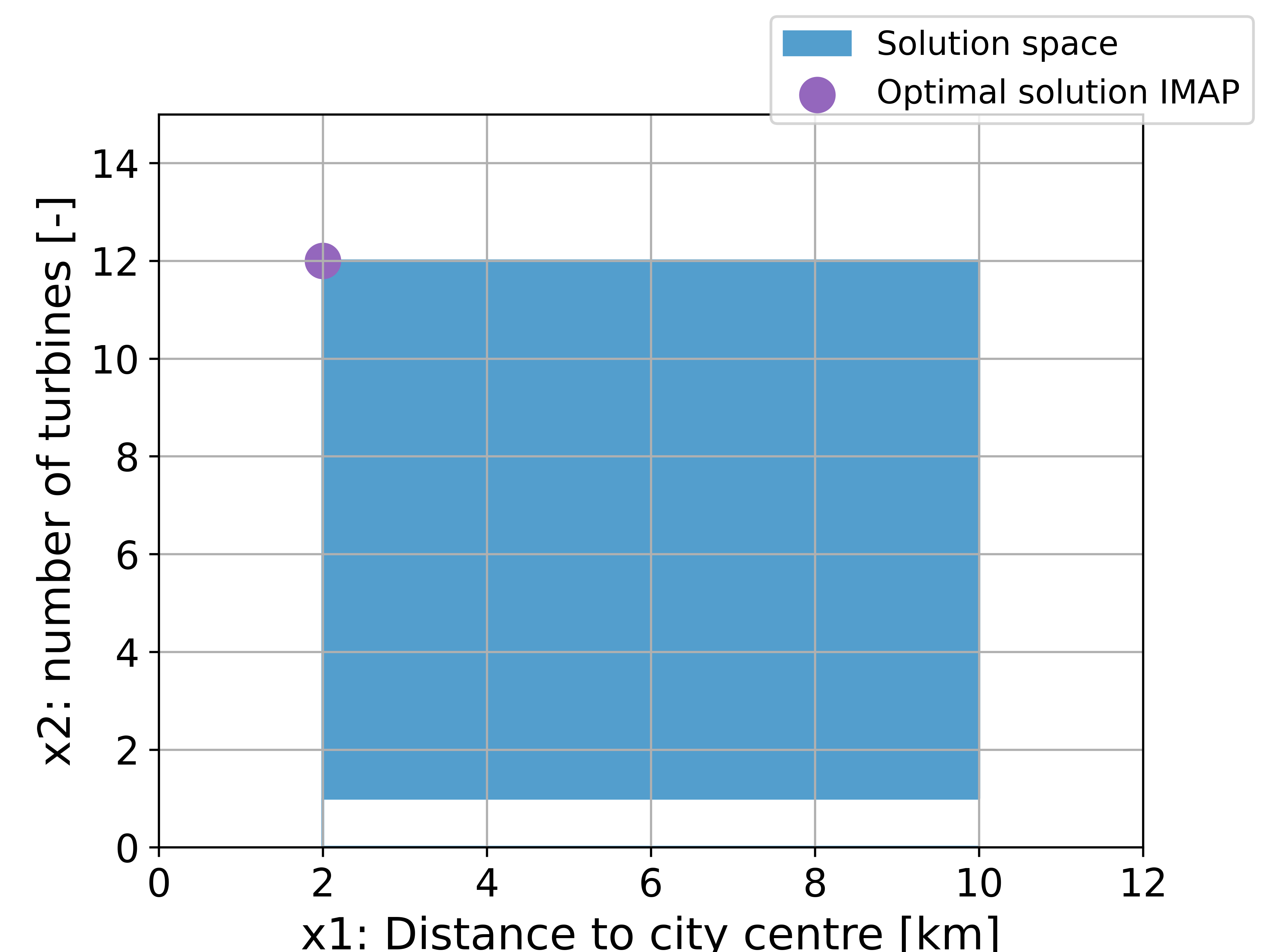}
    \end{minipage}
    \caption{The preference functions including best-fit design point and solution space.}
    \label{fig:2x2_100_energy}
\end{figure}

\subsubsection{100\% S2}\label{2x2_100_residents}

\begin{table}[H]
    \centering
    \caption{A best-fit design point}
    \begin{tabular}{cc}
    \toprule
    Variable     &  IMAP outcome\\
    \midrule
    $x_1$     & 10 km \\
    $x_2$ &  1 \# \\
    \bottomrule
    \end{tabular}
    \label{tab:2x2 local results}
\end{table}

\begin{figure}[H]
    \centering
    \begin{minipage}[b]{0.32\linewidth}
        \includegraphics[width=\linewidth]{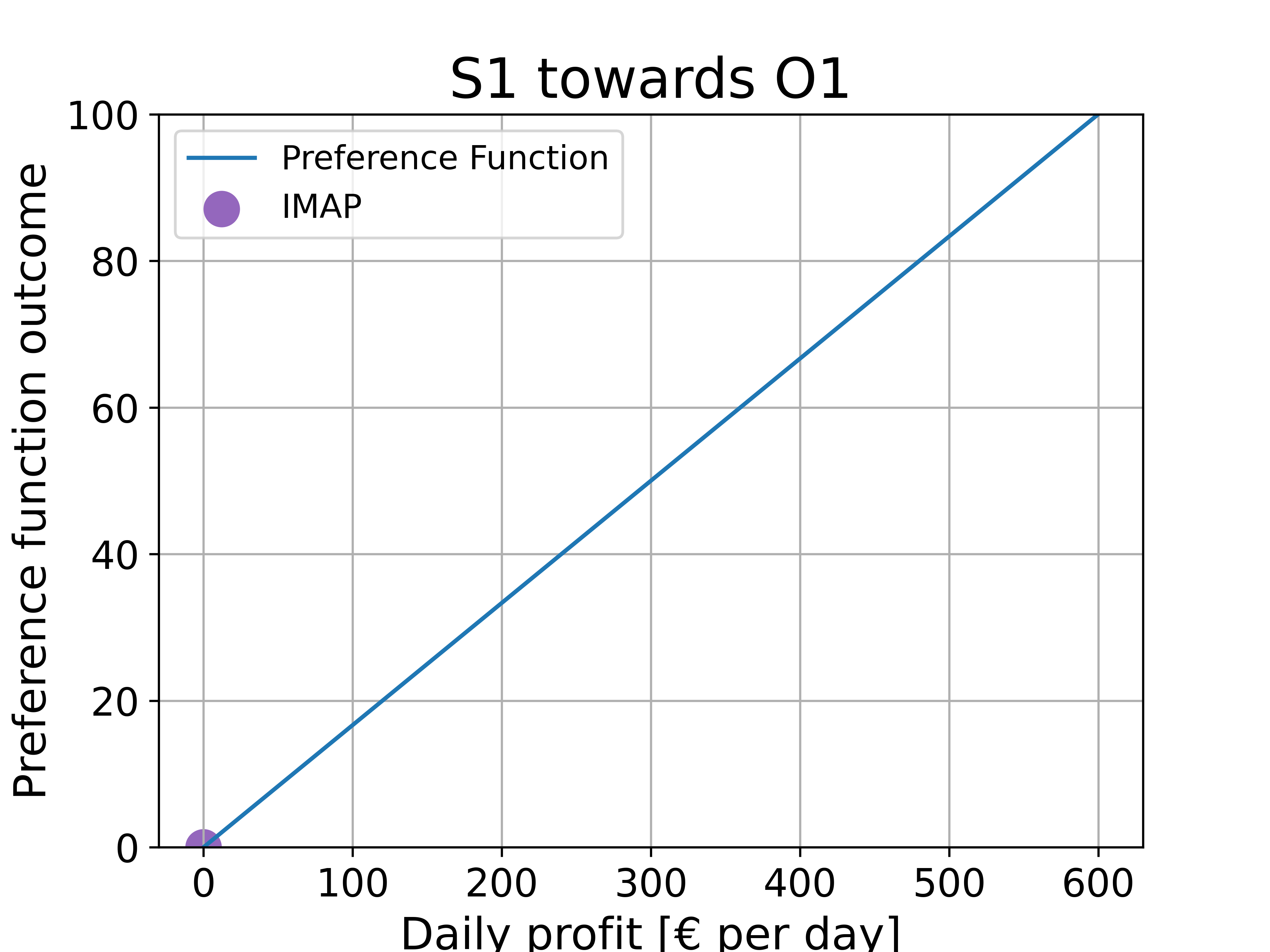}
    \end{minipage}
    \hfill
    \begin{minipage}[b]{0.32\linewidth}
        \includegraphics[width=\linewidth]{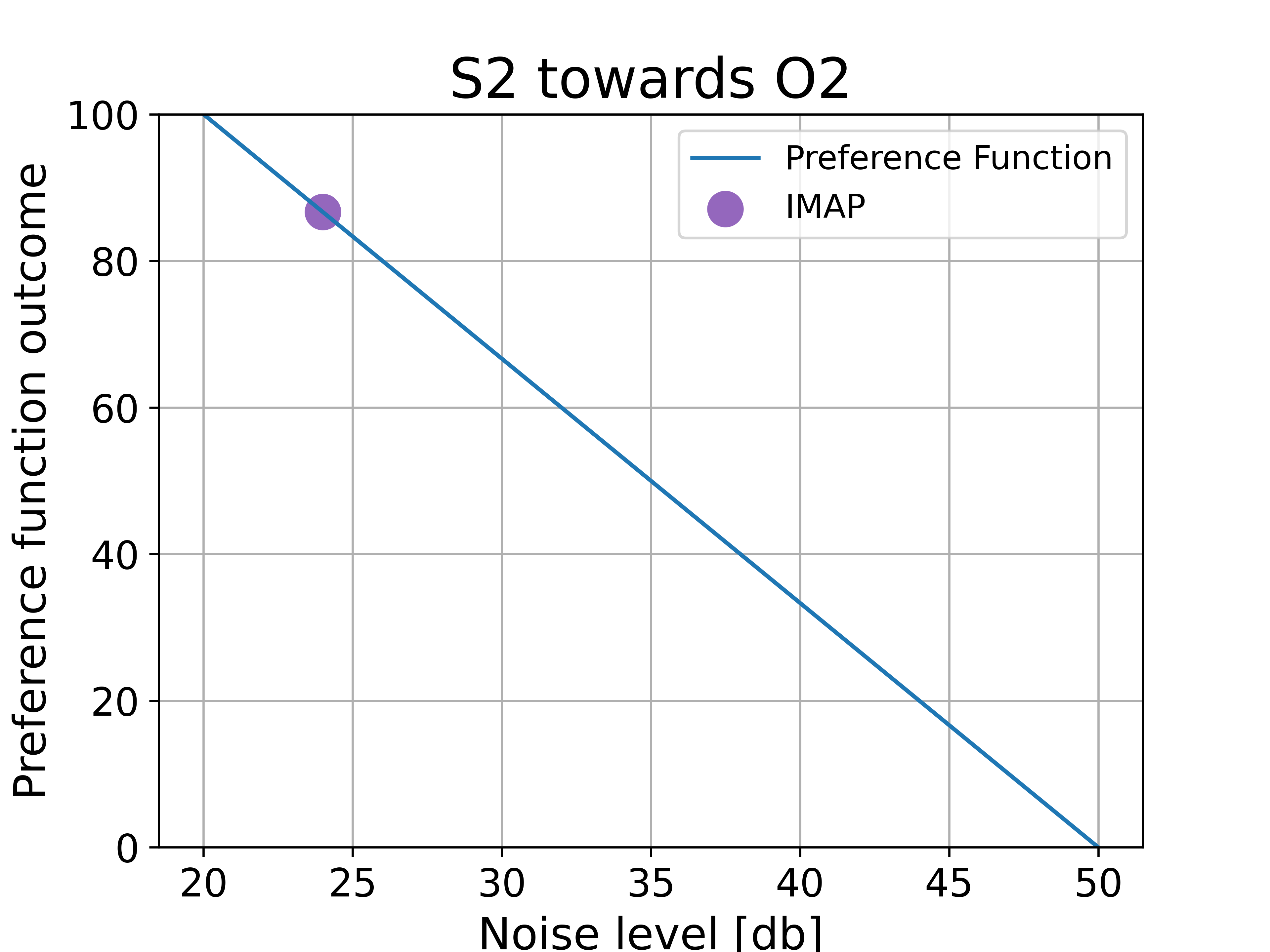}
    \end{minipage}
    \hfill
    \begin{minipage}[b]{0.32\linewidth}
        \includegraphics[width=\linewidth]{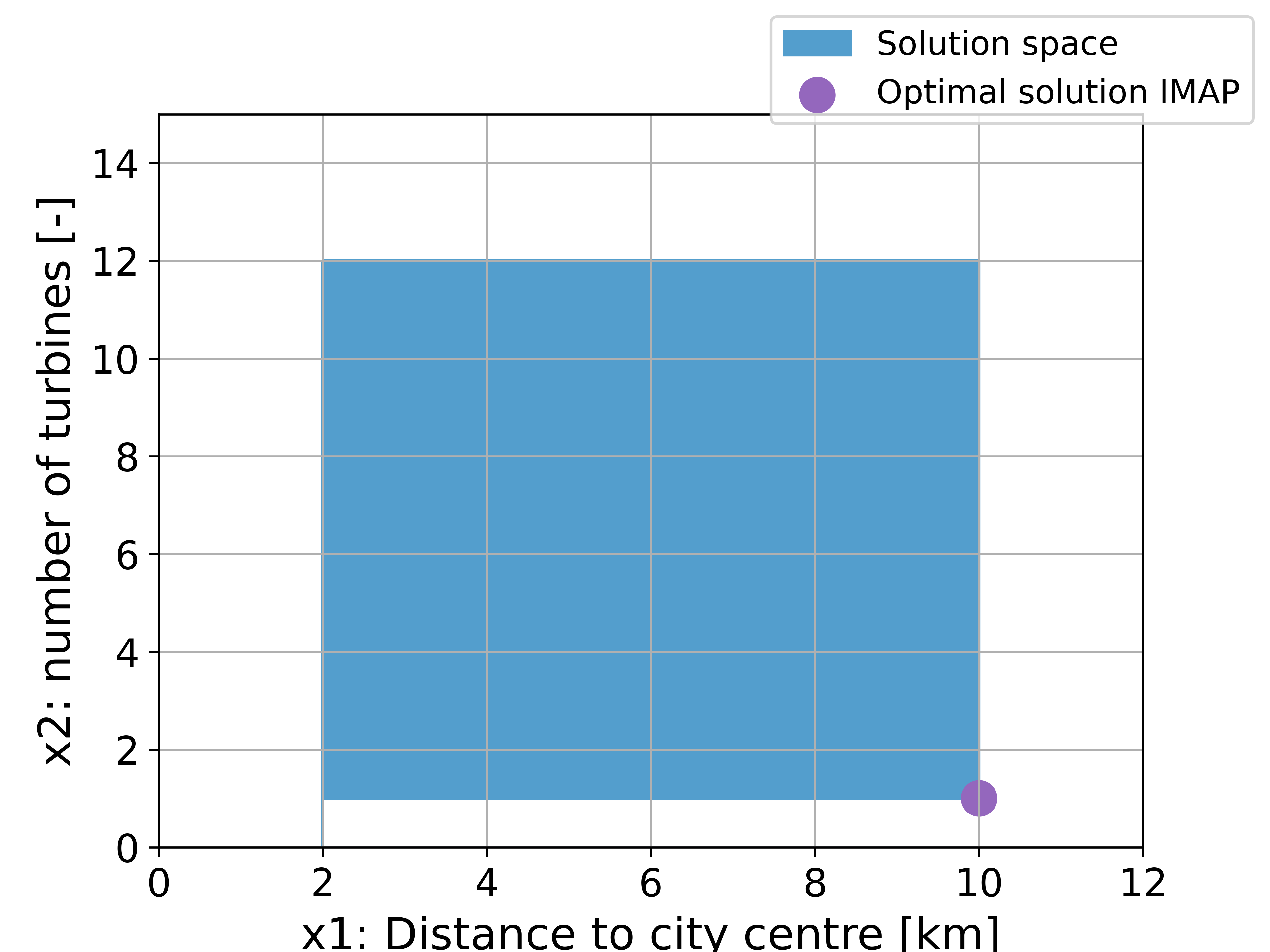}
    \end{minipage}
    \caption{The preference functions including best-fit design point and solution space.}
    \label{fig:2x2_100_local}
\end{figure}

\subsubsection{50/50\% for S1..S2}\label{2x2_50/50}

\begin{table}[H]
    \centering
    \caption{A best-fit design point.}
    \begin{tabular}{cc}
    \toprule
    Variable     &  IMAP outcome\\
    \midrule
    $x_1$     & 10 km \\
    $x_2$ &  12 \#\\
    \bottomrule
    \end{tabular}
    \label{tab:2x2 50 50 results}
\end{table}

\begin{figure}[H]
    \centering
    \begin{minipage}[b]{0.32\linewidth}
        \includegraphics[width=\linewidth]{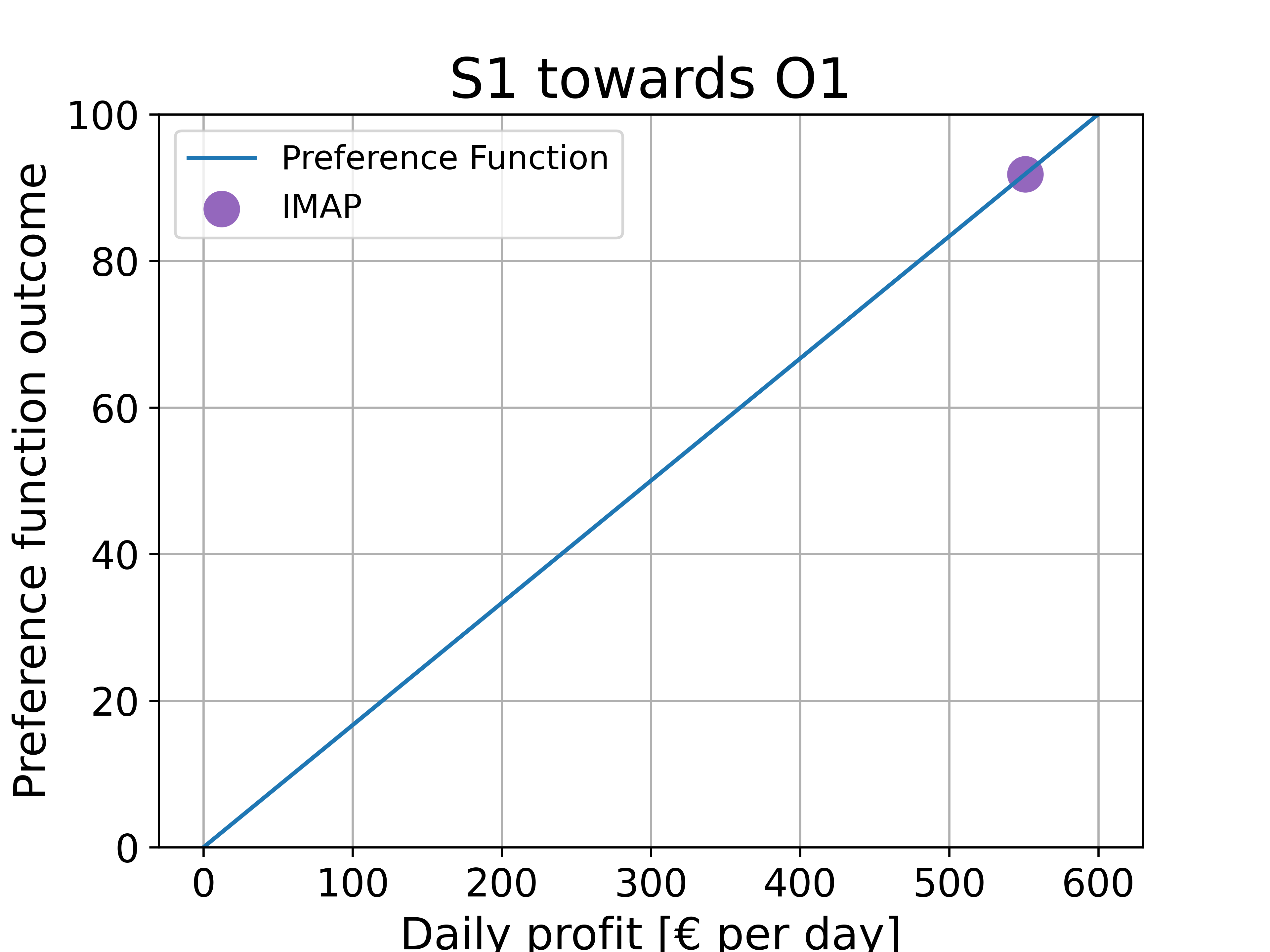}
        \label{fig:2x2_50_50_preference_curve_profit}
    \end{minipage}
    \hfill
    \begin{minipage}[b]{0.32\linewidth}
        \includegraphics[width=\linewidth]{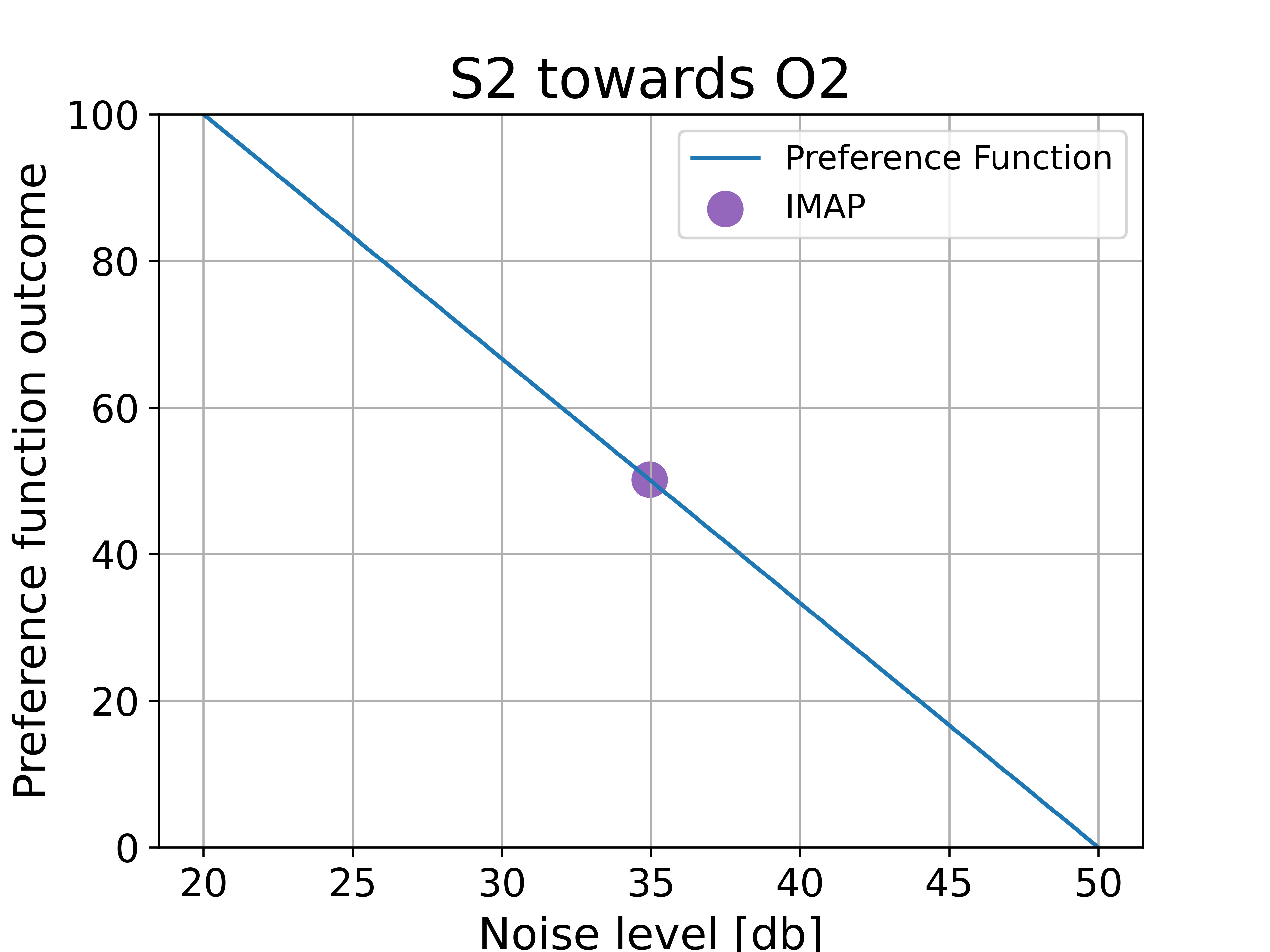}
        \label{fig:2x2_50_50_preference_curve_noise}
    \end{minipage}
    \hfill
    \begin{minipage}[b]{0.32\linewidth}
        \includegraphics[width=\linewidth]{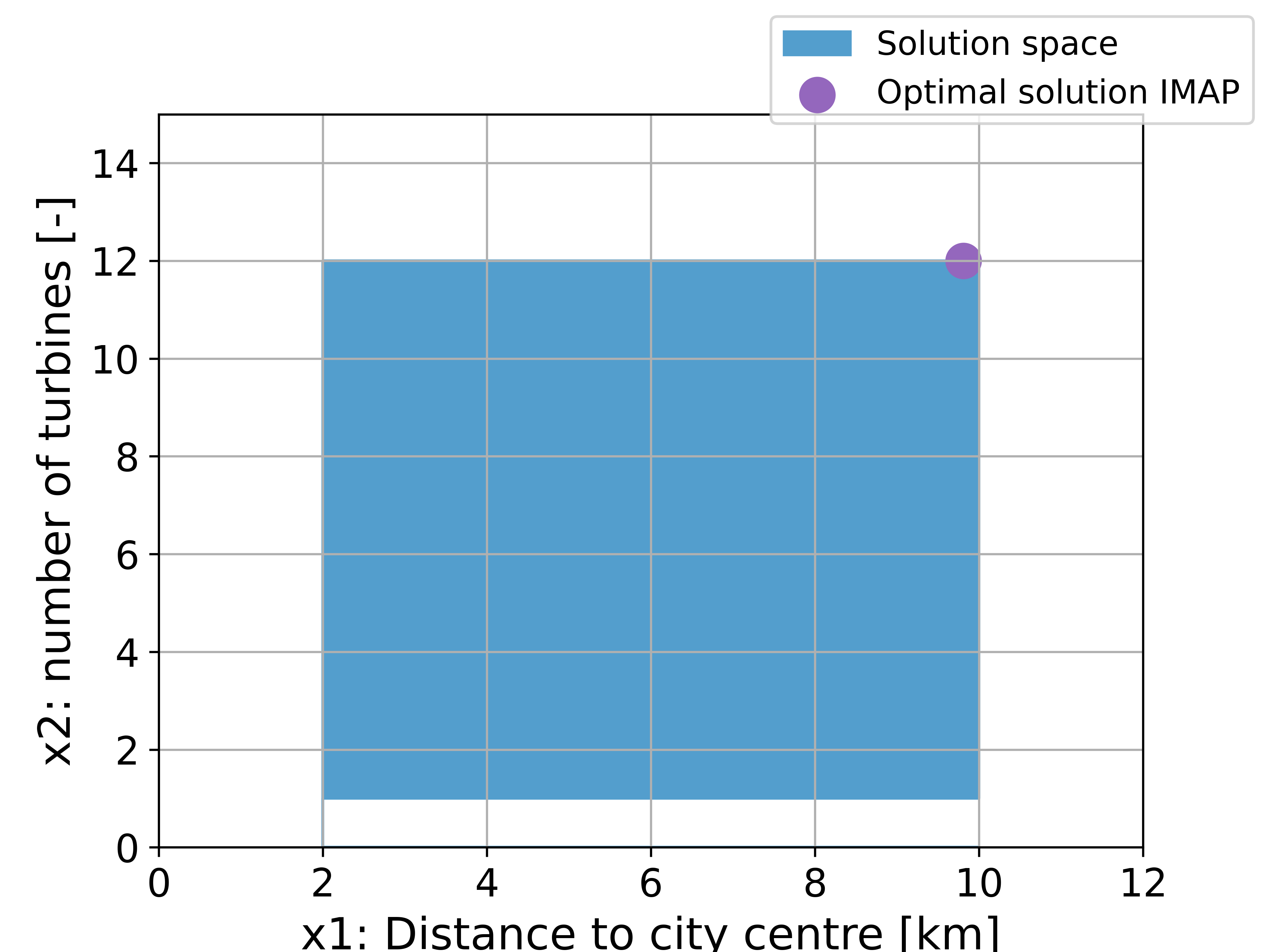}
        \label{fig:2x2_50_50_solution_space}
    \end{minipage}
    \caption{Preference functions including best-fit design point.}
    \label{fig:2x2_50_50}
\end{figure}

\subsection{Case 2: linear 4x4 single-interest}
In these case 2 results, see the quantitative charts and data in the next subsections, we can notice the following high over:
\begin{itemize}
    \item In case 2a, it can be seen that a design synthesis of socio-technical conflicts of interest generated by the Preferendus is most driven by 'desirability'. Because there are 3 stakeholders together ‘against’ and only the energy provider is with 50\% ‘in favour’ of a windfarm development, a best-fit seems viable with balanced preference function results. However, the windspeed required shows that the solution is physically unfeasible.
    \item Case 2b shows that a design synthesis of the socio-technical conflicts of interest design outcomes generated by the Preferendus is most strongly driven by 'capability'. Because the four stakeholders participate equally, the importance of the service provider has declined compared to 2a, making the best-fit see a non-viable wind farm and the design point falls on the endpoints of the preference functions (completely 'satisfied' or 'dissatisfied').
\end{itemize}

\subsubsection{Uneven 50/20/15/15\% for S1..S4}

\begin{table}[H]
    \centering
    \caption{A best-fit design point}
    \begin{tabular}{ll}
    \toprule
    Variable     &  IMAP outcome\\
    \midrule
    $x_1$     & 10 km \\
    $x_2$ &  11 \# \\
    $x_3$ &  140 m\\
    $x_4$ & 14.8 m/s\\
    \bottomrule
    \end{tabular}
    \label{tab:2x2 local results}
\end{table}

\begin{figure}[H]
    \centering
    \begin{minipage}[b]{0.40\linewidth}
        \includegraphics[width=\linewidth]{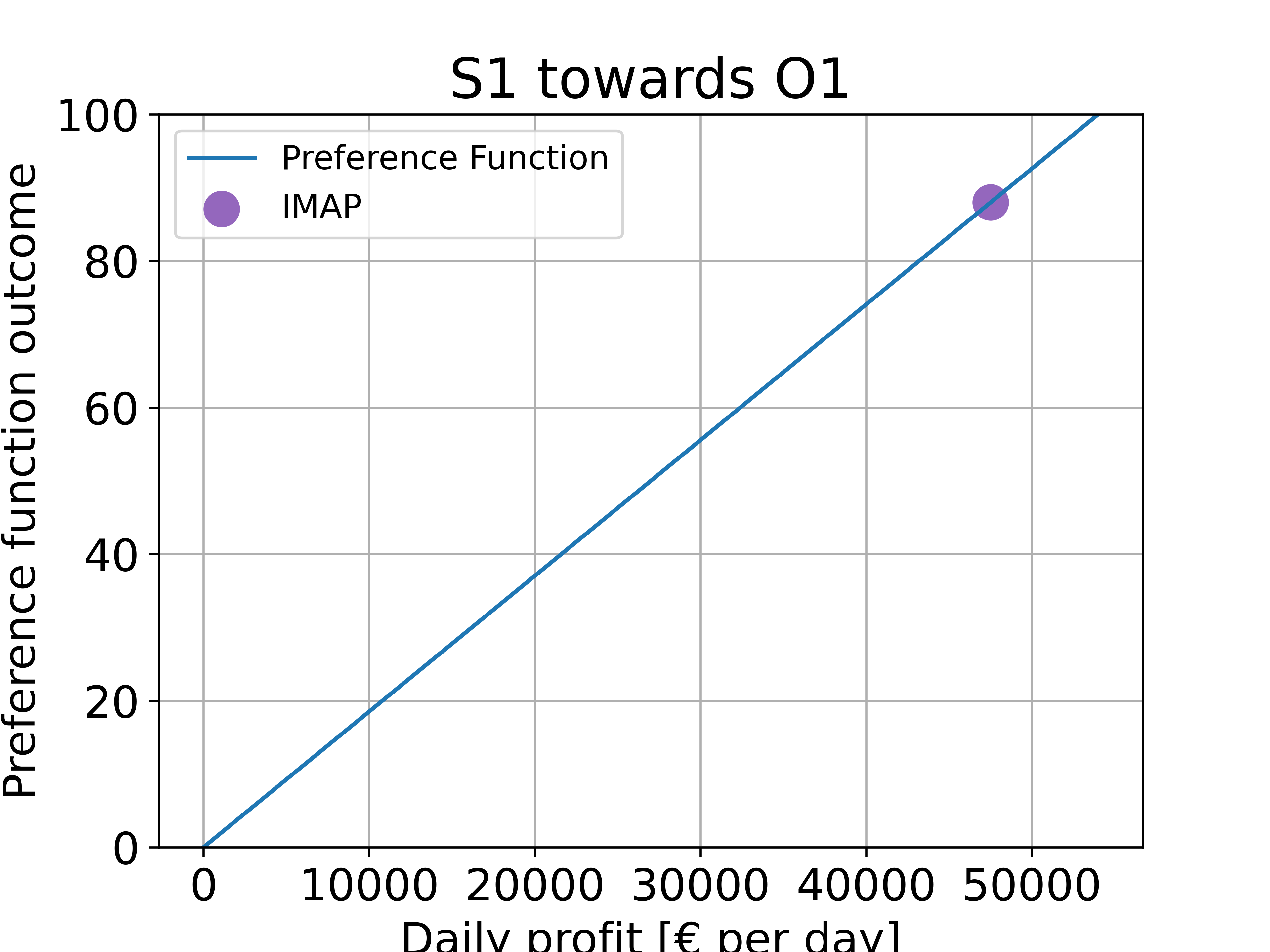}
    \end{minipage}
    \hfill
    \begin{minipage}[b]{0.40\linewidth}
        \includegraphics[width=\linewidth]{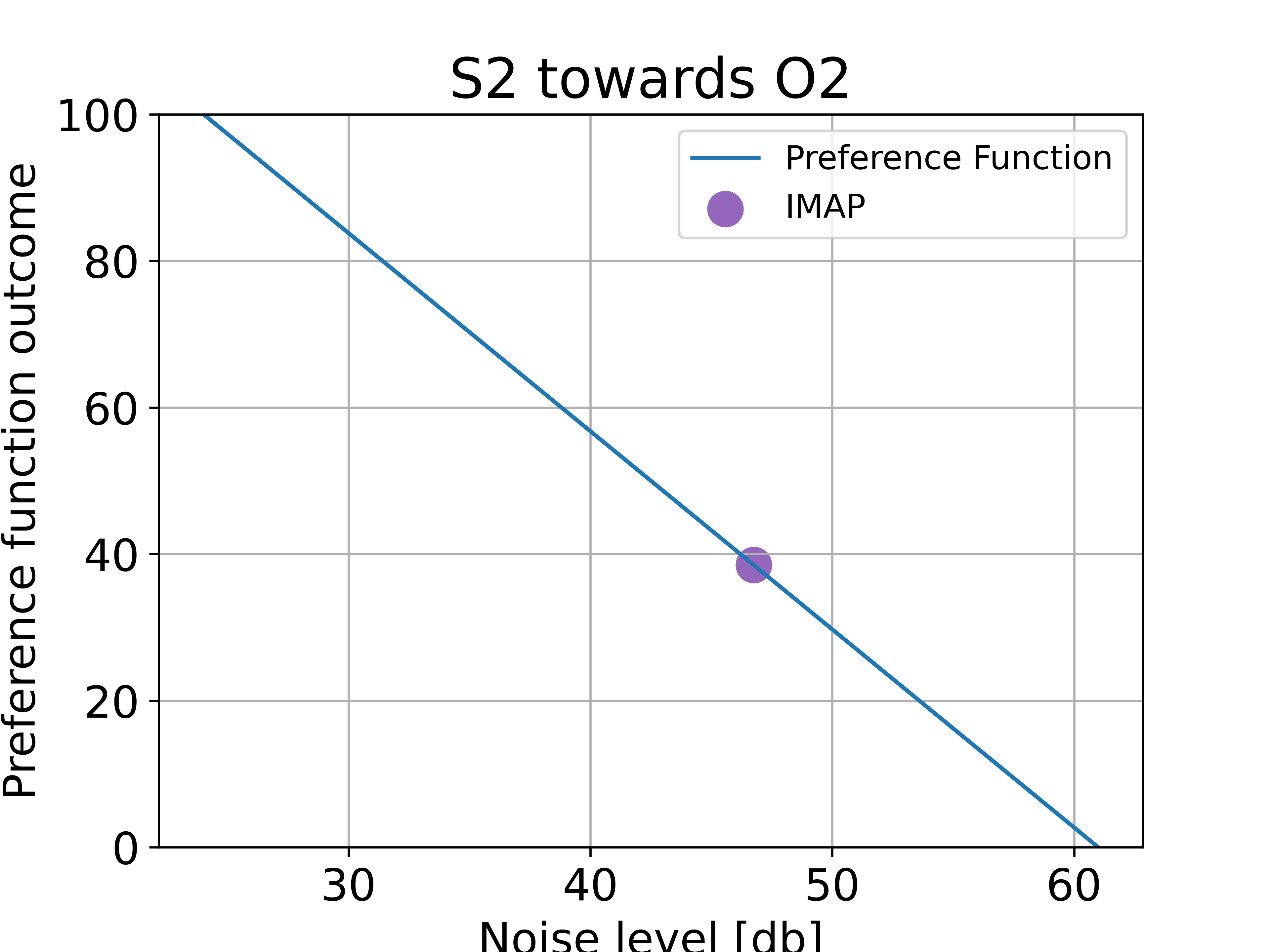}
    \end{minipage}
    
    \vspace{0.5cm} 

    \begin{minipage}[b]{0.40\linewidth}
        \includegraphics[width=\linewidth]{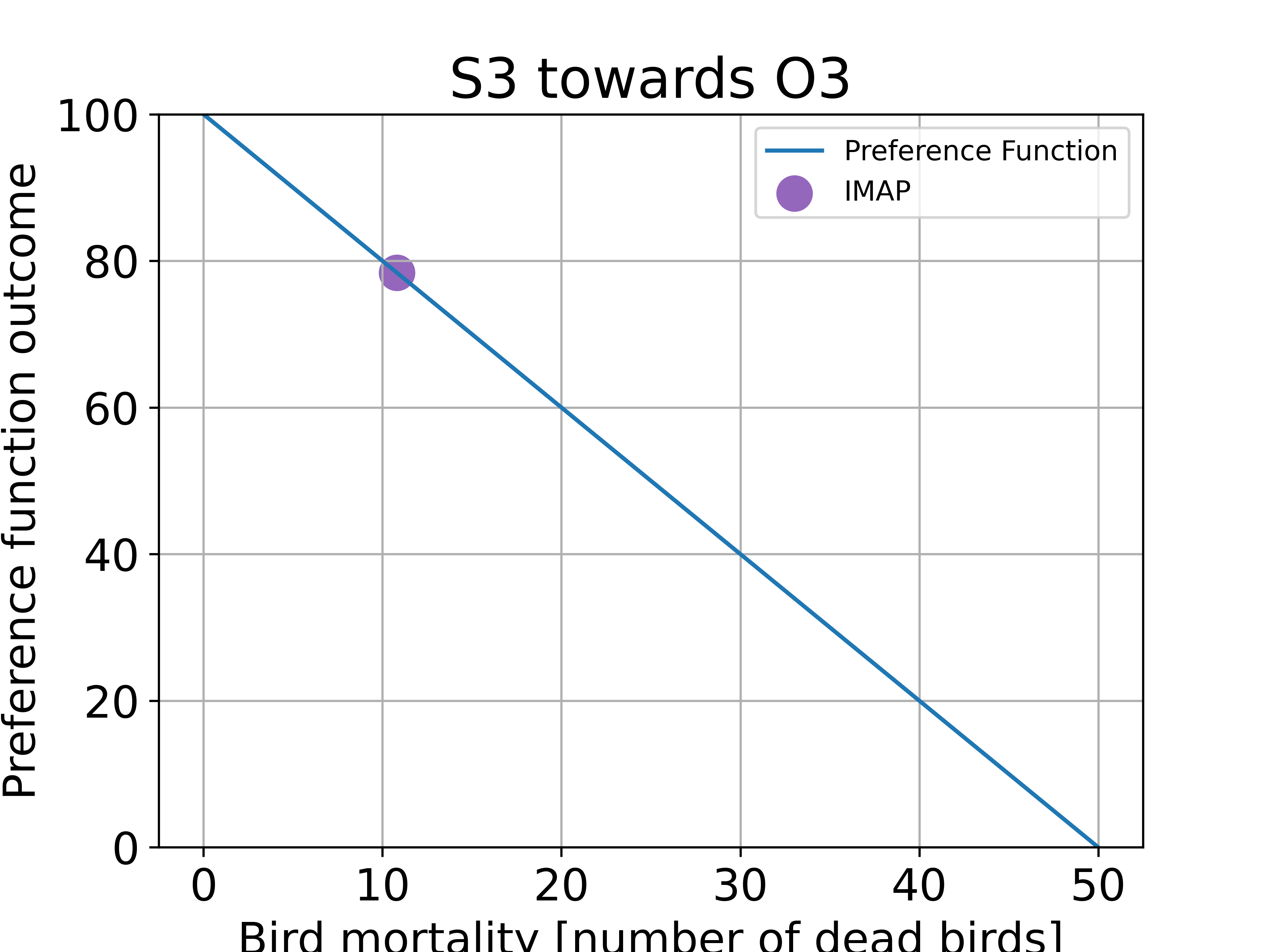}
    \end{minipage}
    \hfill
    \begin{minipage}[b]{0.40\linewidth}
        \includegraphics[width=\linewidth]{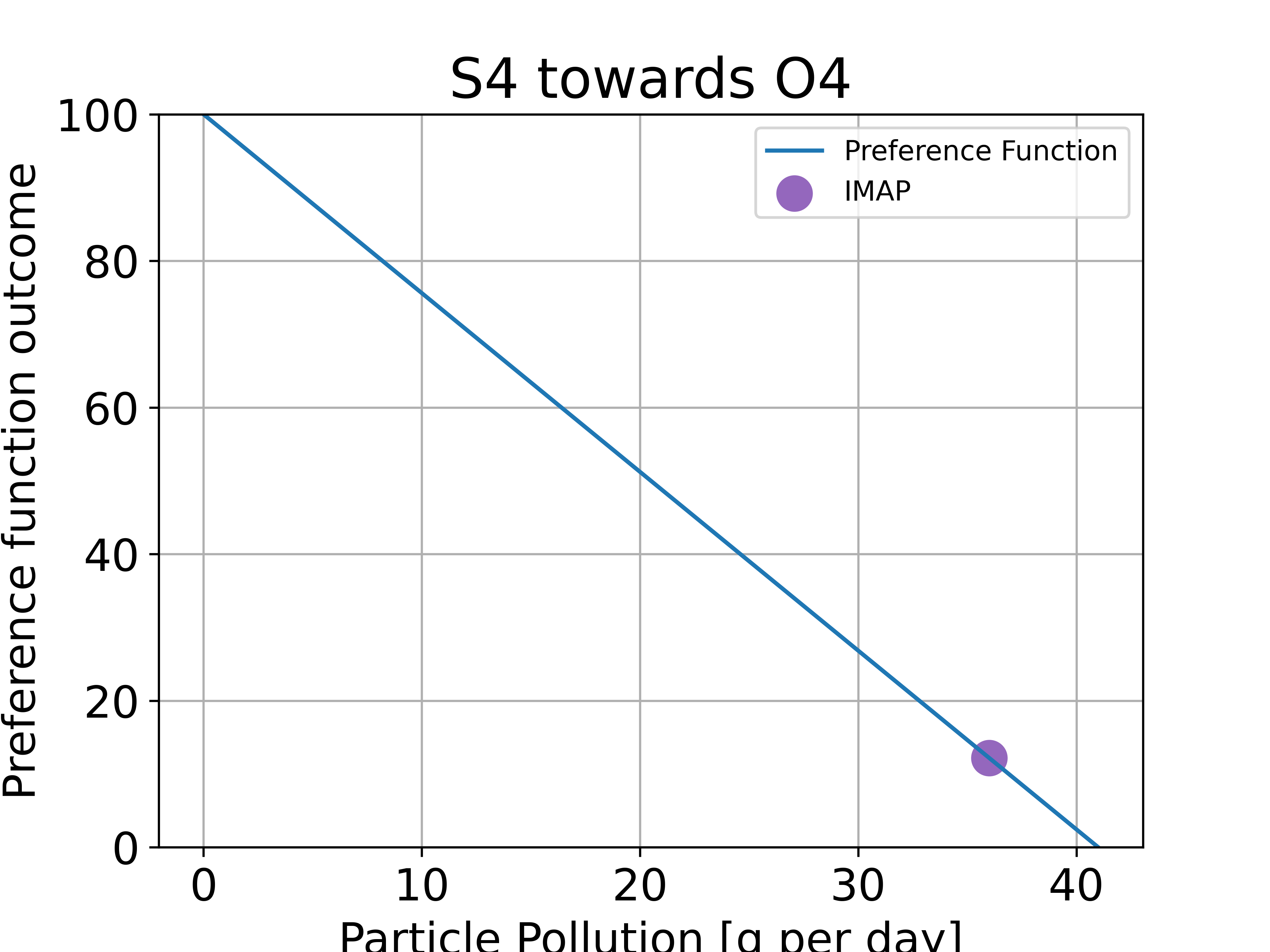}
    \end{minipage}
    \caption{Preference functions including best-fit design point.}
    \label{fig:4x4_energy}
\end{figure}

\subsubsection{Equal 25/25/25/25\% for S1..S4}

\begin{table}[H]
    \centering
    \caption{A best-fit design point.}
    \begin{tabular}{ll}
    \toprule
    Variable     &  IMAP outcome\\
    \midrule
    $x_1$     & 10 km \\
    $x_2$ &  1 \# \\
    $x_3$ & 50 m \\
    $x_4$ & 3.0 m/s\\
    \bottomrule
    \end{tabular}
    \label{tab:4x4 equal results}
\end{table}

\begin{figure}[H]
    \centering
    \begin{minipage}[b]{0.40\linewidth}
        \includegraphics[width=\linewidth]{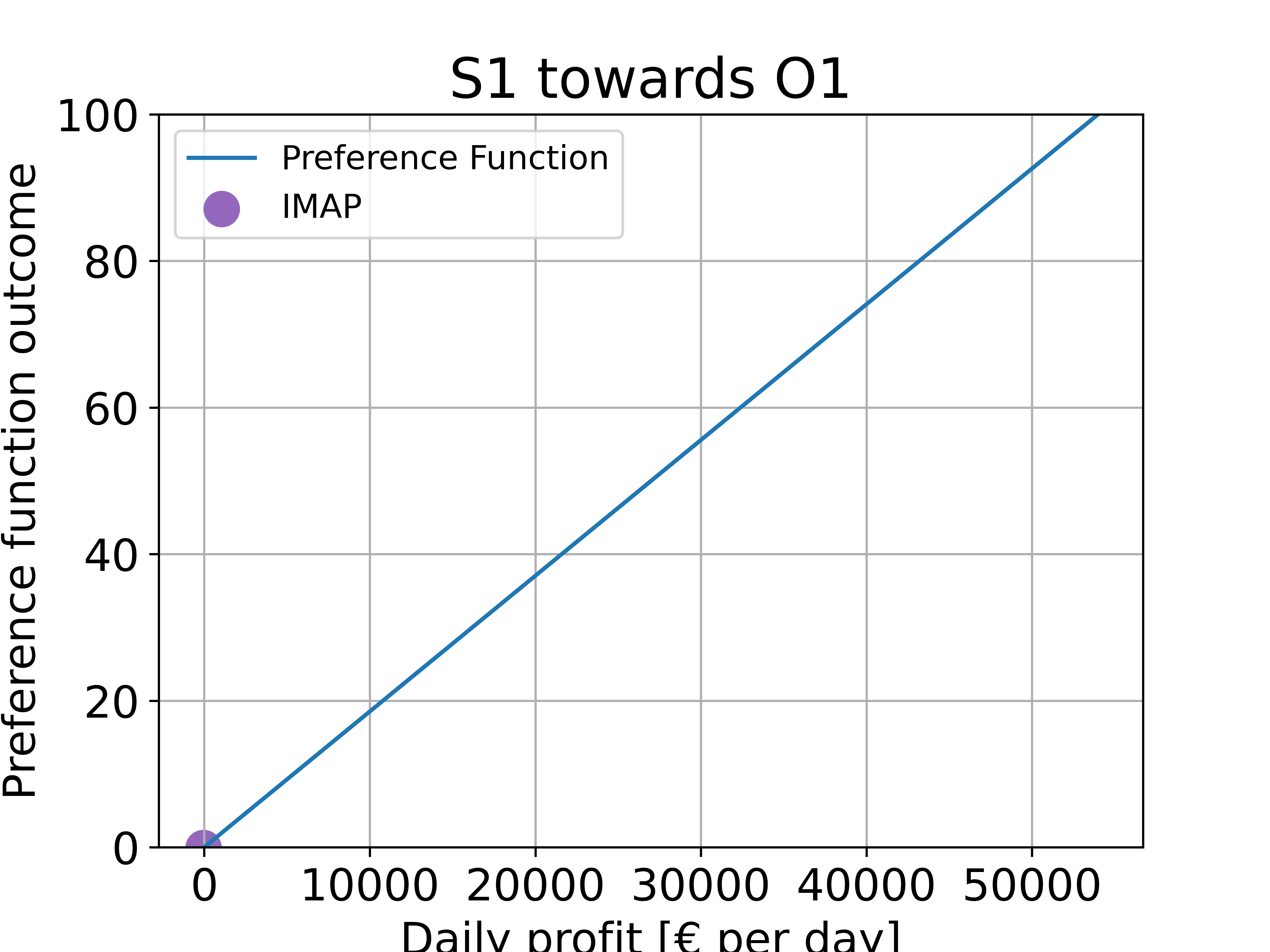}
    \end{minipage}
    \hfill
    \begin{minipage}[b]{0.40\linewidth}
        \includegraphics[width=\linewidth]{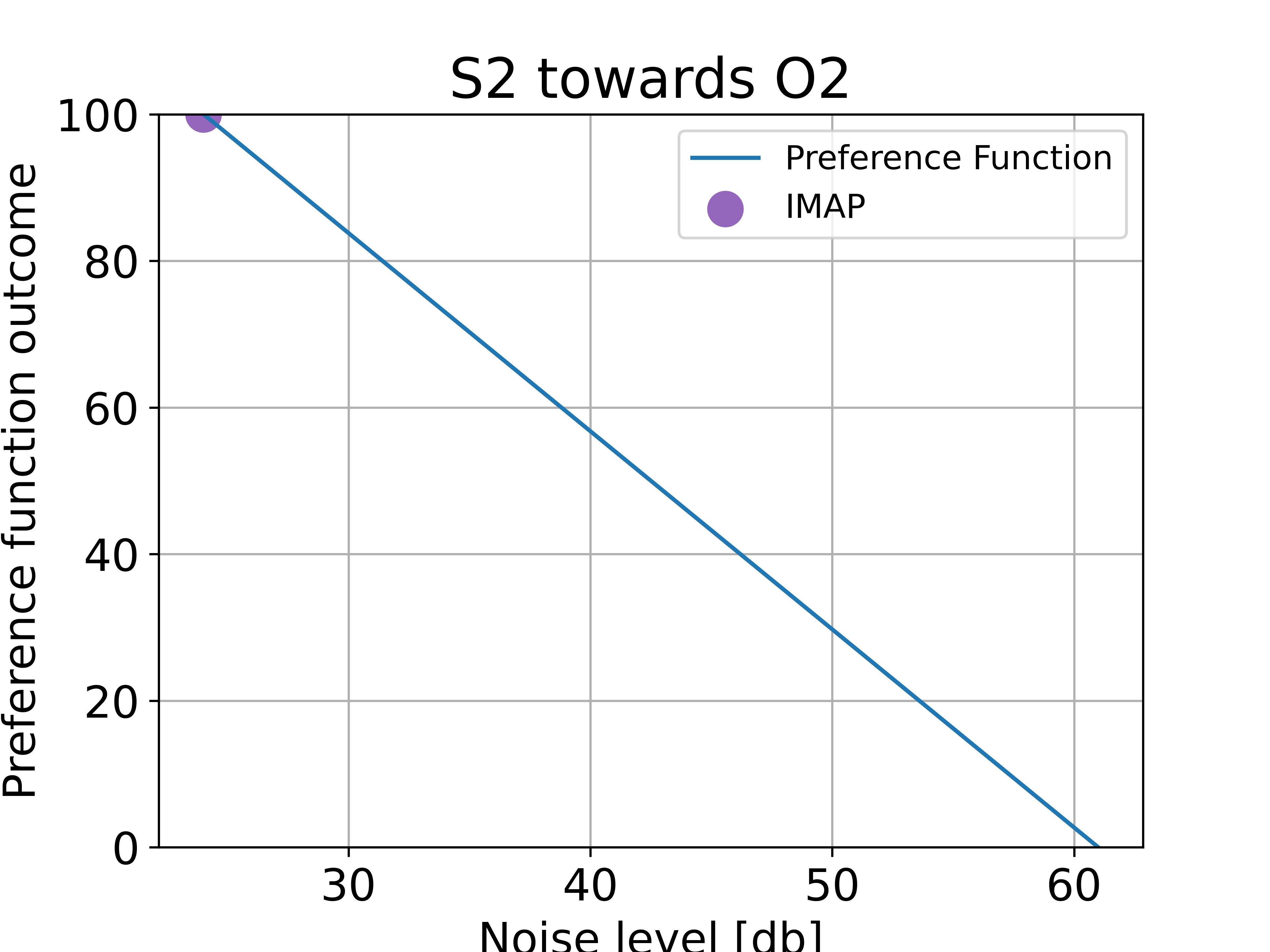}
    \end{minipage}
    
    \vspace{0.5cm} 

    \begin{minipage}[b]{0.40\linewidth}
        \includegraphics[width=\linewidth]{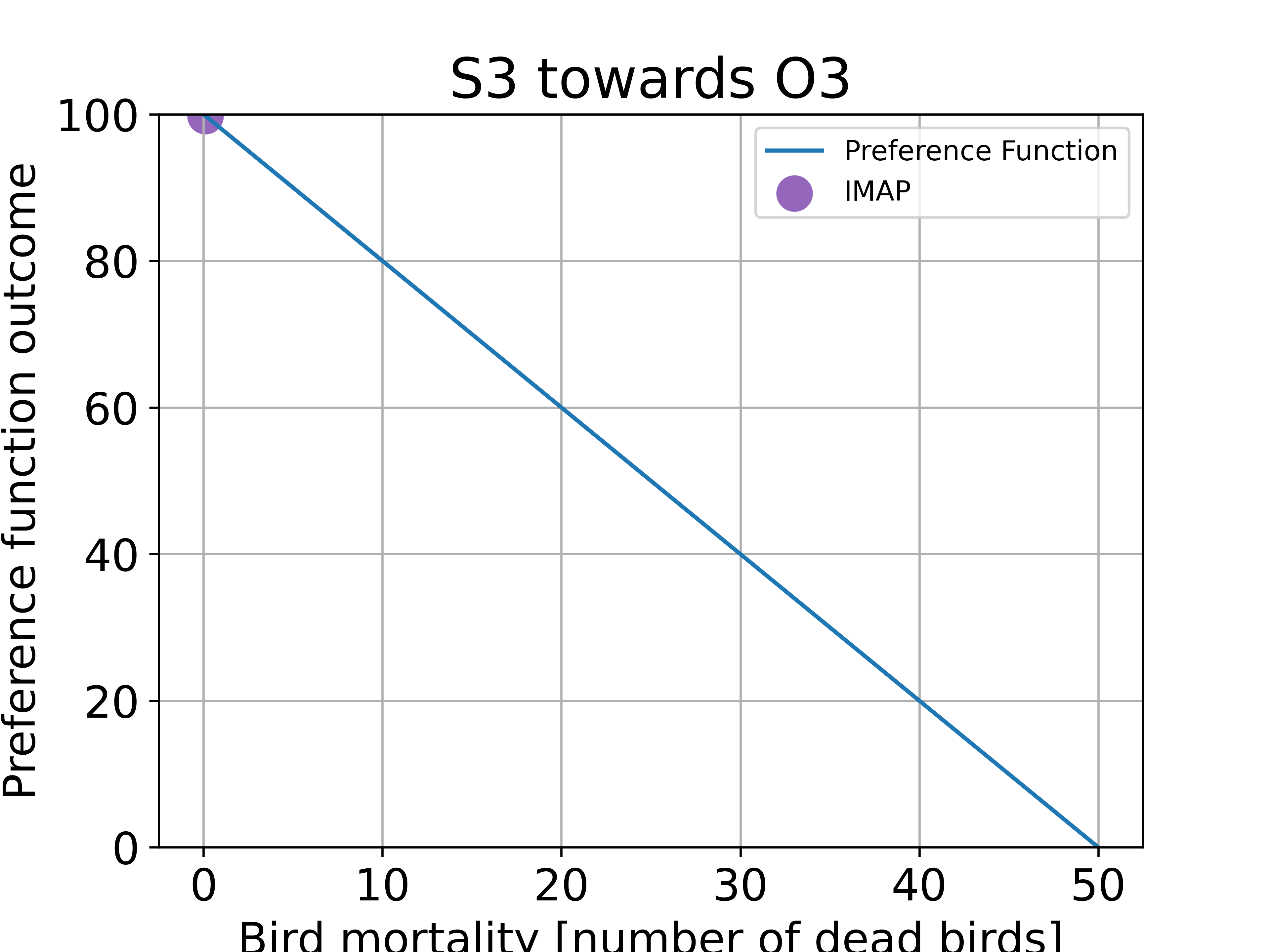}
    \end{minipage}
    \hfill
    \begin{minipage}[b]{0.40\linewidth}
        \includegraphics[width=\linewidth]{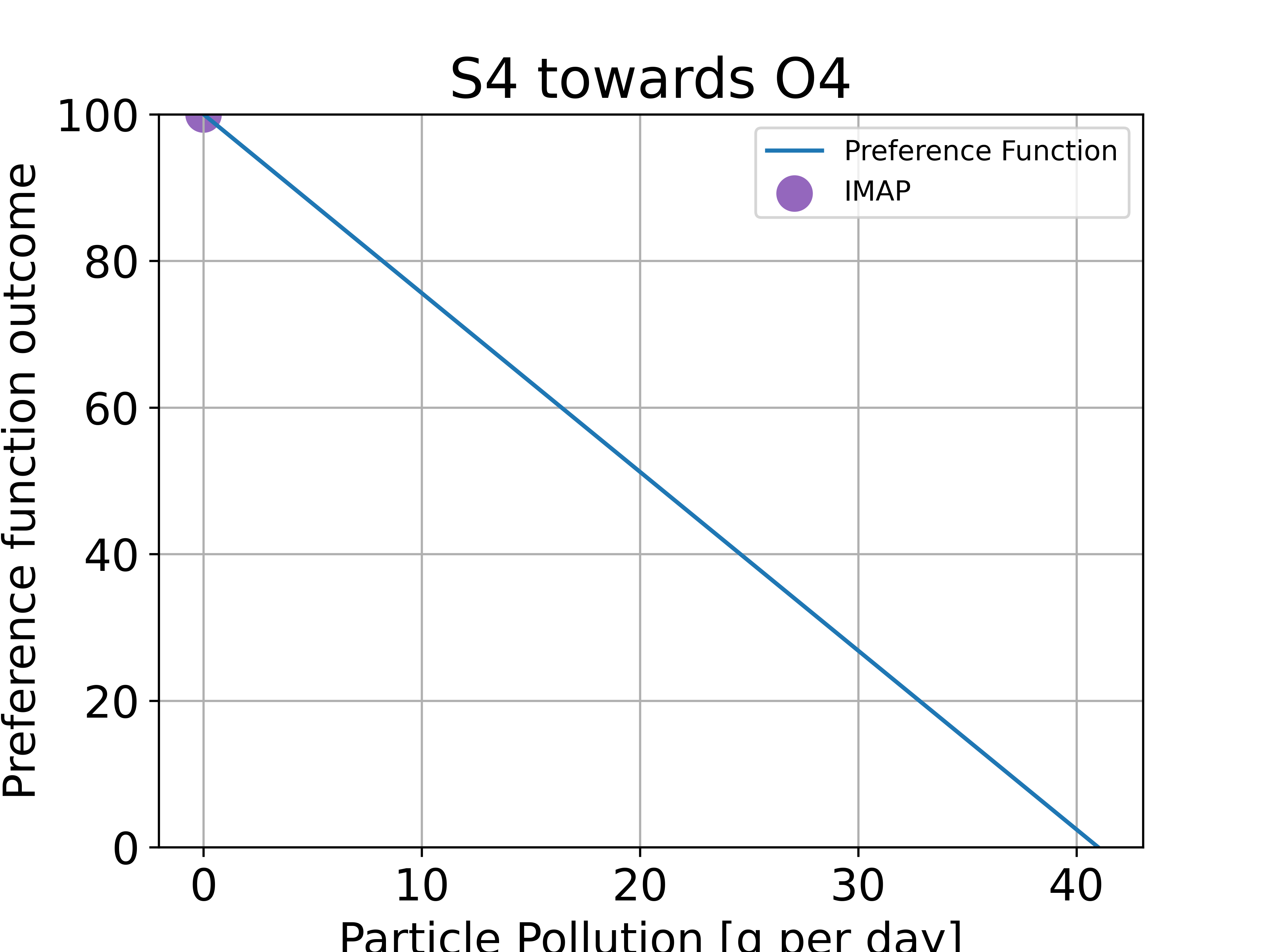}
    \end{minipage}
    \caption{Preference functions including best-fit design point.}
    \label{fig:4x4_equal}
\end{figure}

\subsection{Case 3: non-linear 4x4 multiple-interests}
In these case 3 results, see the quantitative charts hereafter, we can notice the following high over :

\begin{itemize}
    \item Case 3 shows that a design synthesis of socio-technical conflicts of interest generated by the Preferendus is driven most strongly by ‘desirability’, in that there is a ‘coalition’ that seems to be in favour of wind energy revenues ('we will always be better off for that seems to be the credo' and/or 'the desire of this thought seems greater than the reality and its side effect'). However, the required wind speed shows that the solution is actually physically unfeasible.
    \item Case 3, due to the distribution of individual weights and the different preference function shapes, shows a balanced outcome where mostly the design point generated by the Preferendus lies on the function rather than on an endpoint. This case shows similarities with case 2a, partly because the summed weights per objective are rather similar to the global (and uneven) distribution in case 2a.
  
\end{itemize}

\begin{table}[H]
    \centering
    \caption{Solution of 4x4 case where the energy provider has a weight of 50\% “against” the coalition of the other stakeholders.}
    \begin{tabular}{ll}
    \toprule
    Variable     &  IMAP outcome\\
    \midrule
    $x_1$     & 10 km \\
    $x_2$ &  10 \#\\
    $x_3$ &  145 m\\
    $x_4$ & 15.0 m/s \\
    \bottomrule
    \end{tabular}
    \label{tab:2x2 local results}
\end{table}

\begin{figure}[H]
    \centering
    \begin{minipage}[b]{0.40\linewidth}
        \includegraphics[width=\linewidth]{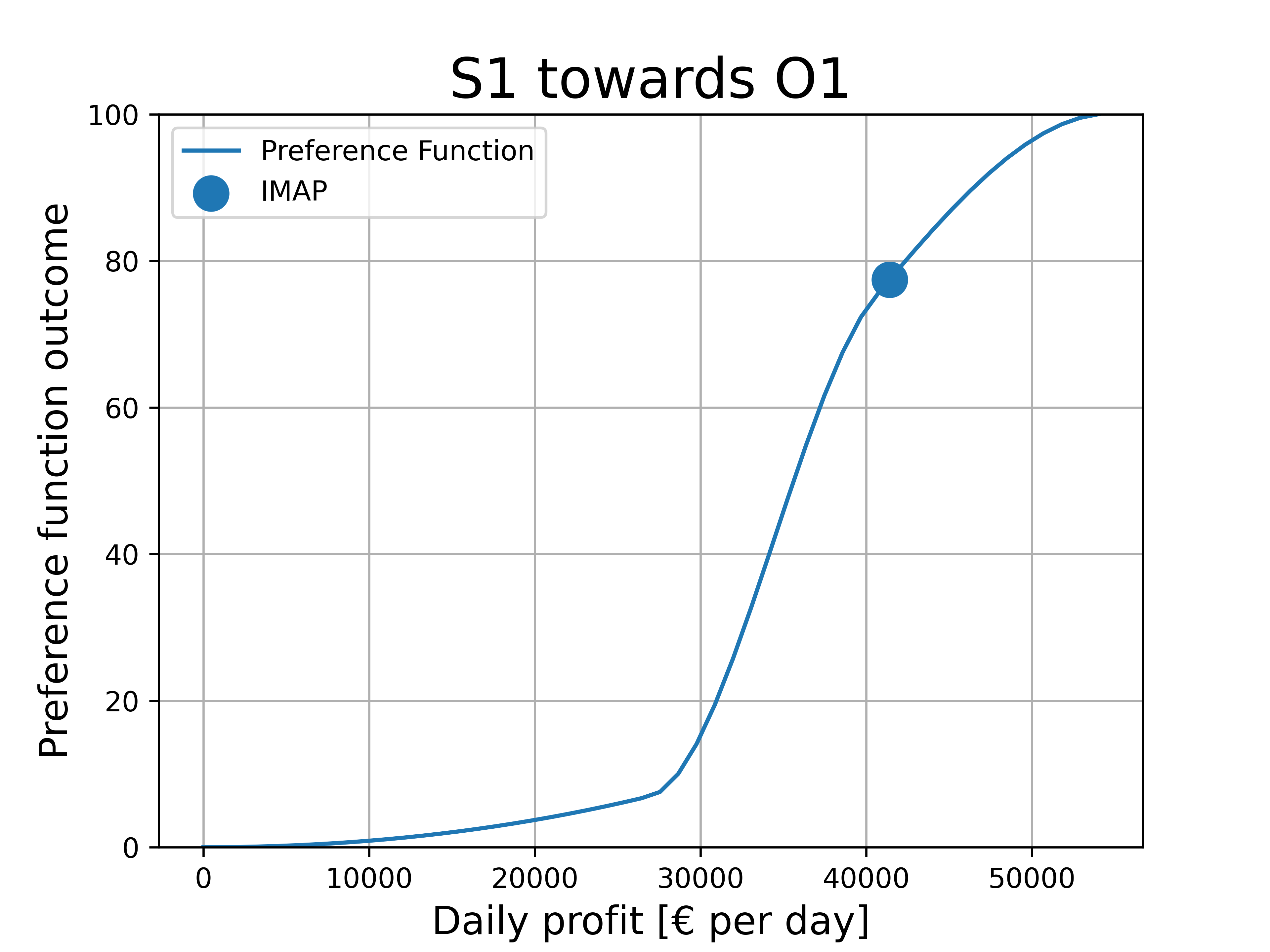}
    \end{minipage}
    \hfill
    \begin{minipage}[b]{0.40\linewidth}
        \includegraphics[width=\linewidth]{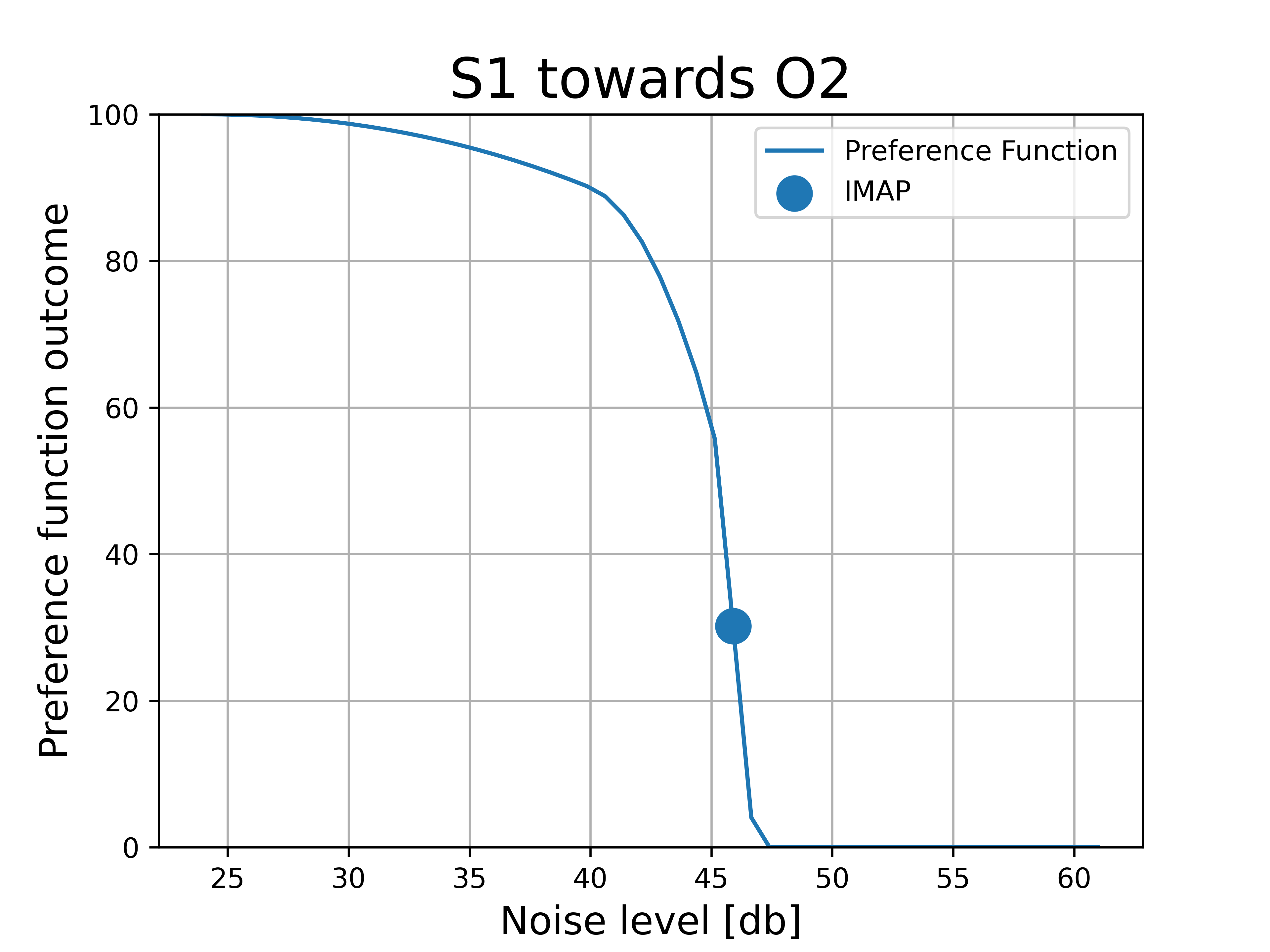}
    \end{minipage}
    
    \vspace{0.5cm} 

    \begin{minipage}[b]{0.40\linewidth}
        \includegraphics[width=\linewidth]{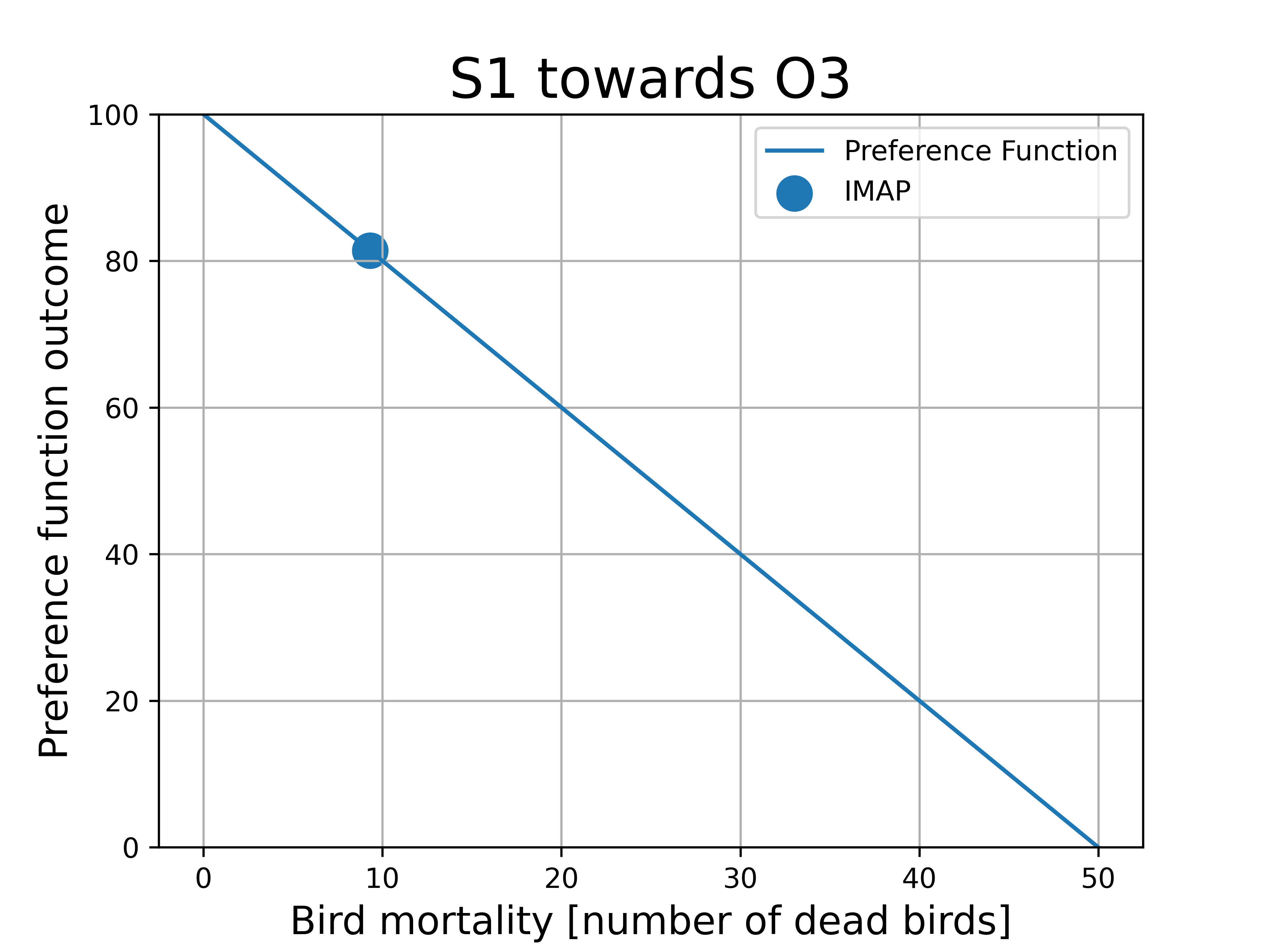}
    \end{minipage}
    \hfill
    \begin{minipage}[b]{0.40\linewidth}
        \includegraphics[width=\linewidth]{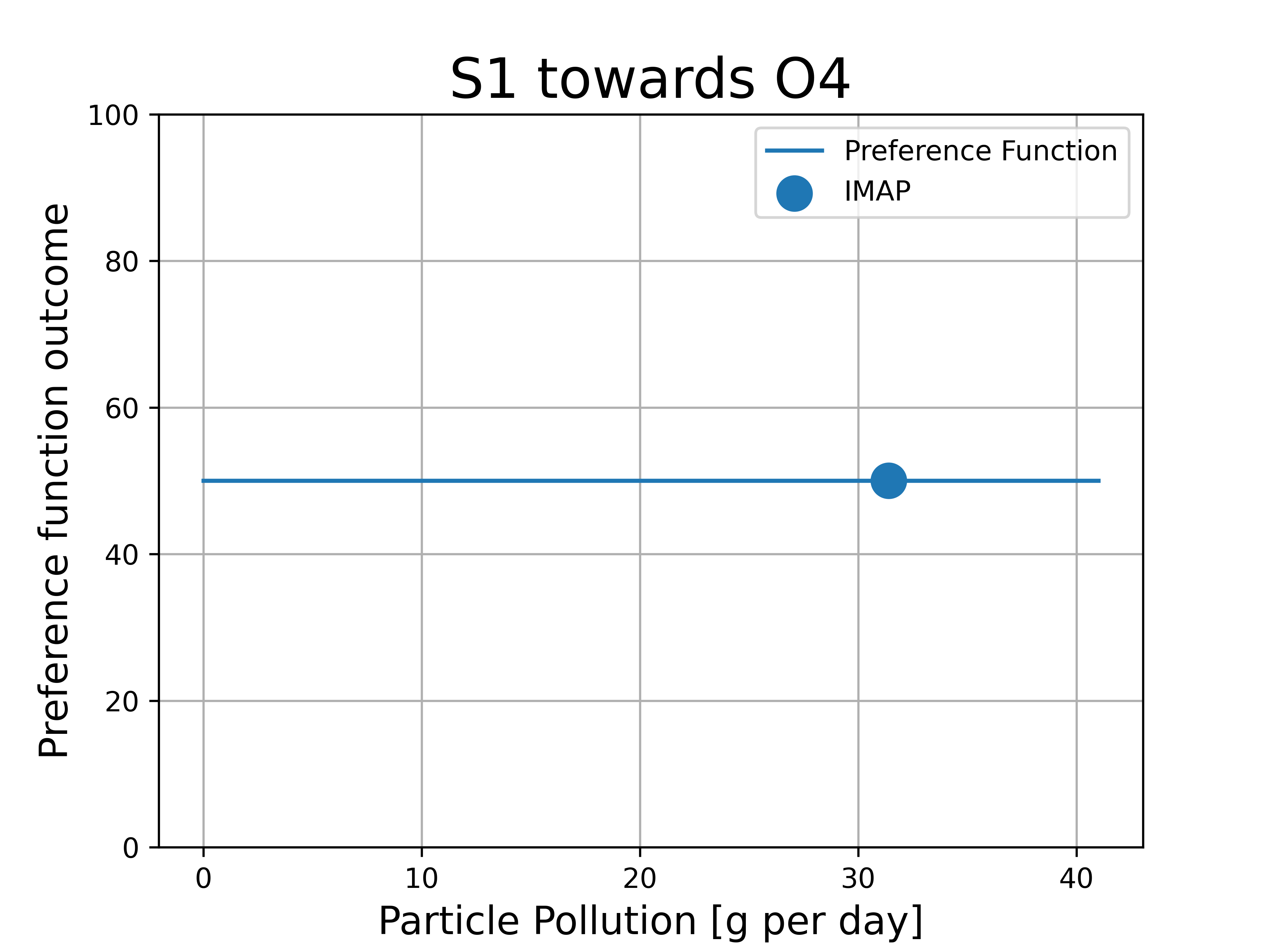}
    \end{minipage}
    \caption{Preference curves for $S_1$ towards $O{1..4}$.}
    \label{fig:4x4_S1}
\end{figure}

\begin{figure}[H]
    \centering
    \begin{minipage}[b]{0.40\linewidth}
        \includegraphics[width=\linewidth]{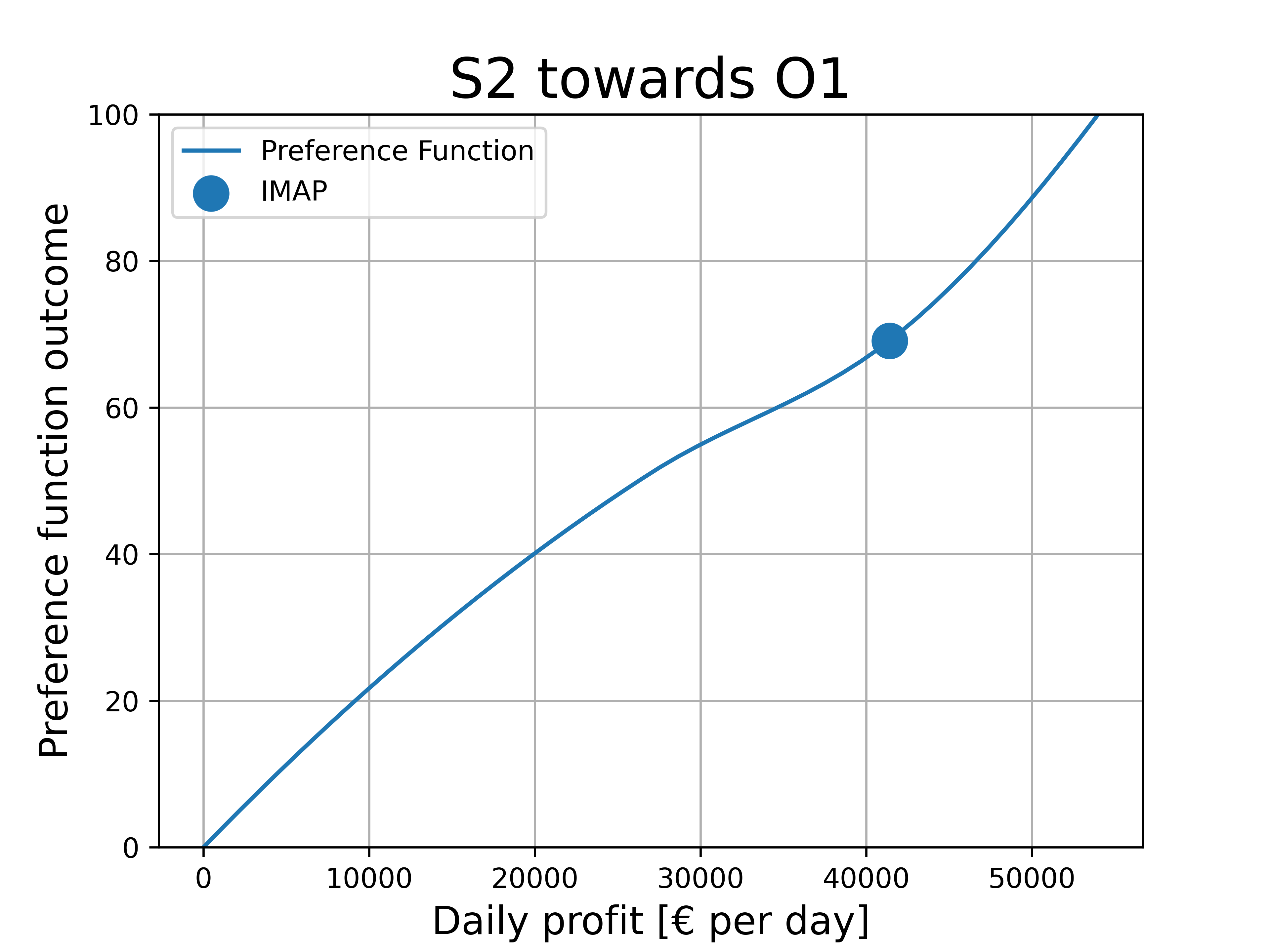}
    \end{minipage}
    \hfill
    \begin{minipage}[b]{0.40\linewidth}
        \includegraphics[width=\linewidth]{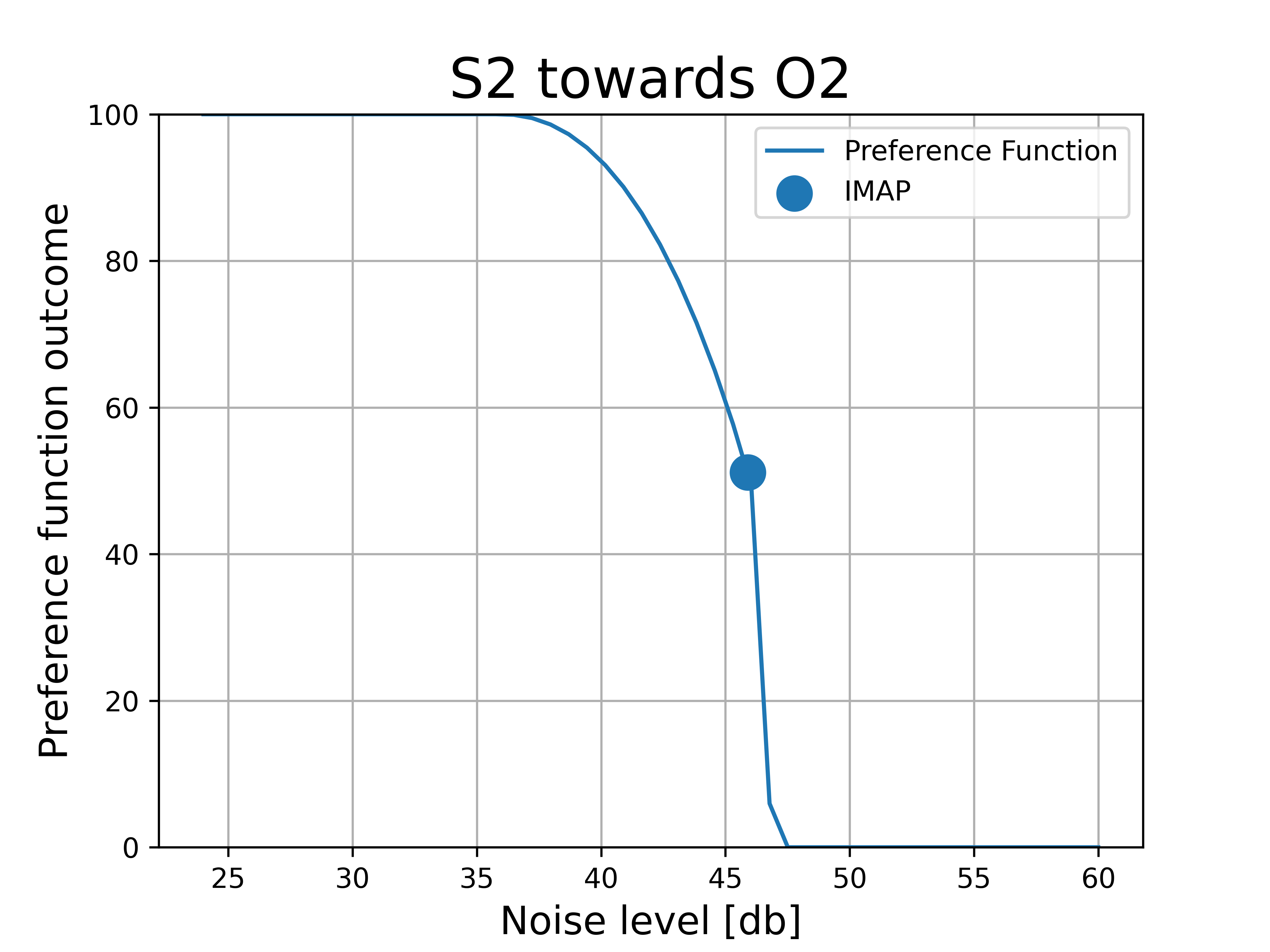}
    \end{minipage}
    
    \vspace{0.5cm} 

    \begin{minipage}[b]{0.40\linewidth}
        \includegraphics[width=\linewidth]{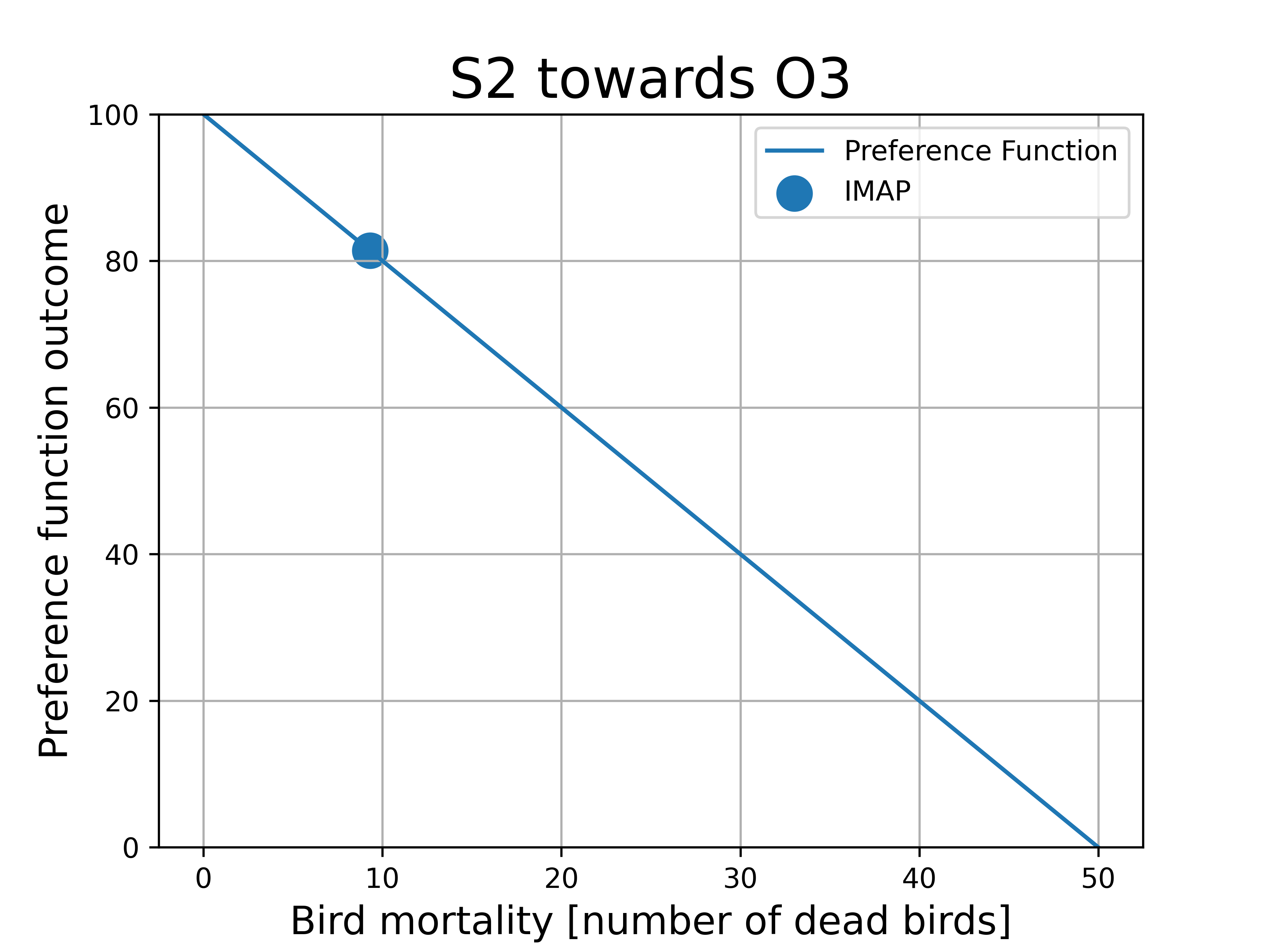}
    \end{minipage}
    \hfill
    \begin{minipage}[b]{0.40\linewidth}
        \includegraphics[width=\linewidth]{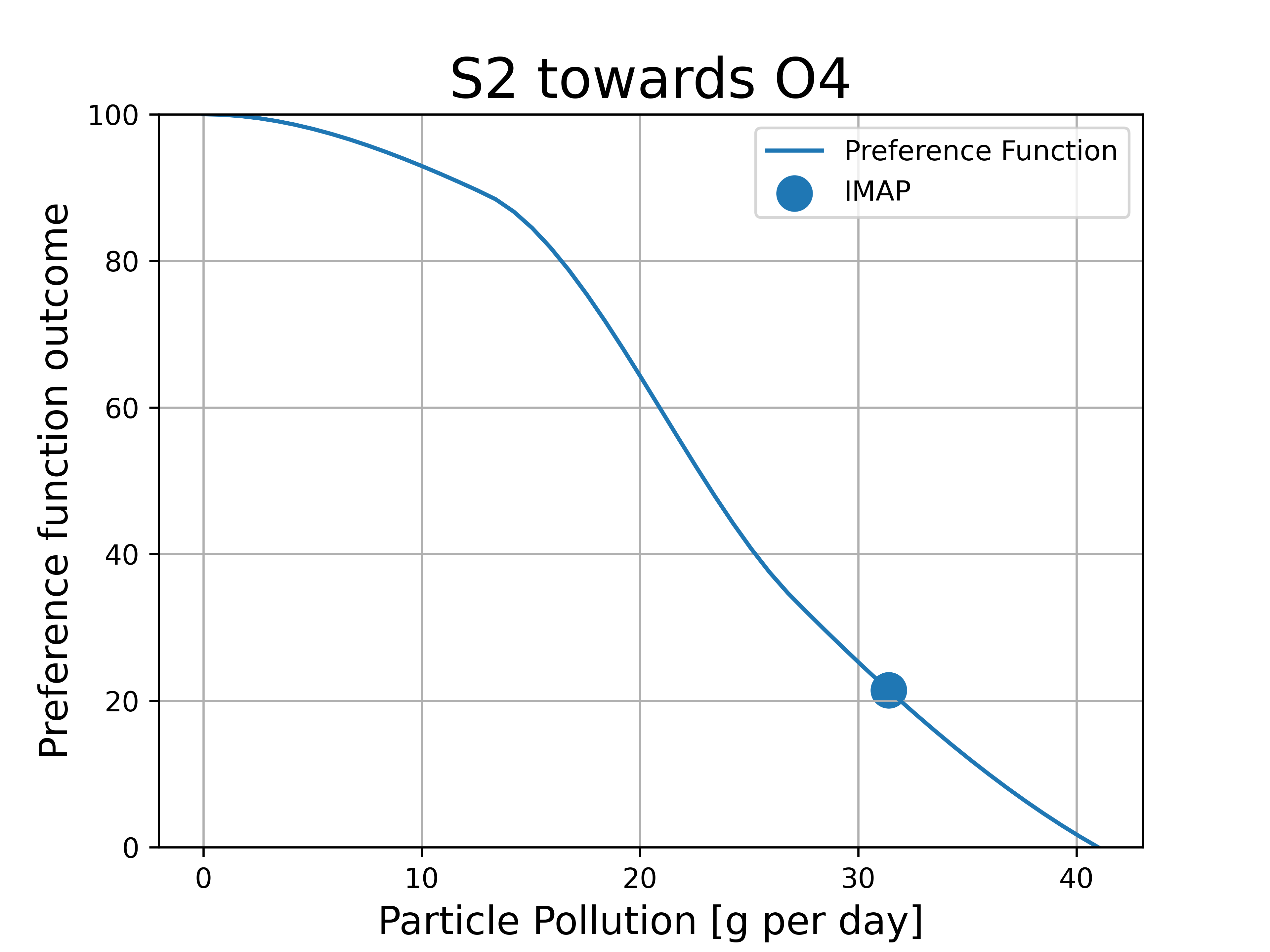}
    \end{minipage}
    \caption{Preference curves for $S_2$ towards $O{1..4}$.}
    \label{fig:4x4_S2}
\end{figure}

\begin{figure}[H]
    \centering
    \begin{minipage}[b]{0.40\linewidth}
        \includegraphics[width=\linewidth]{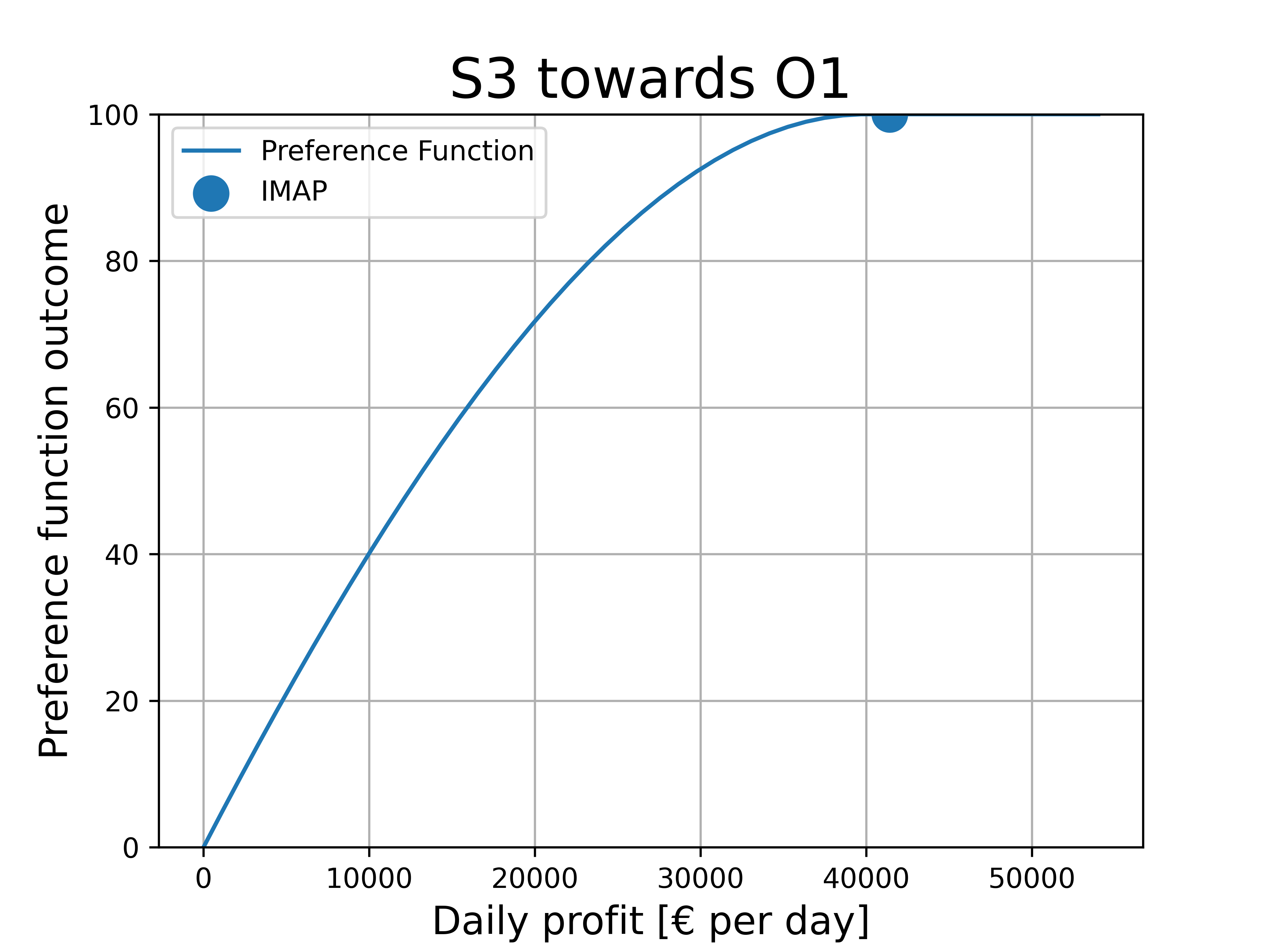}
    \end{minipage}
    \hfill
    \begin{minipage}[b]{0.40\linewidth}
        \includegraphics[width=\linewidth]{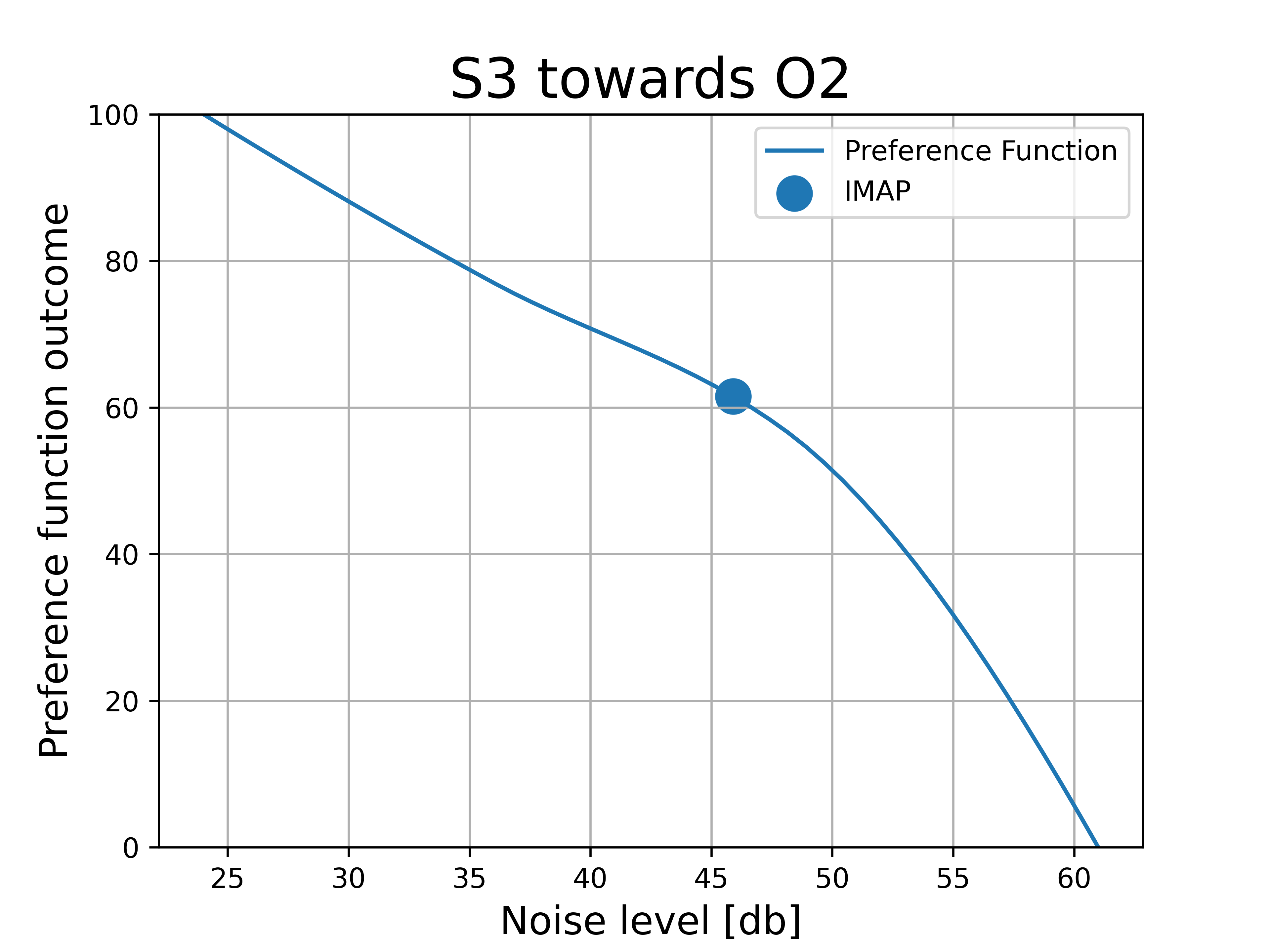}
    \end{minipage}
    
    \vspace{0.5cm} 

    \begin{minipage}[b]{0.40\linewidth}
        \includegraphics[width=\linewidth]{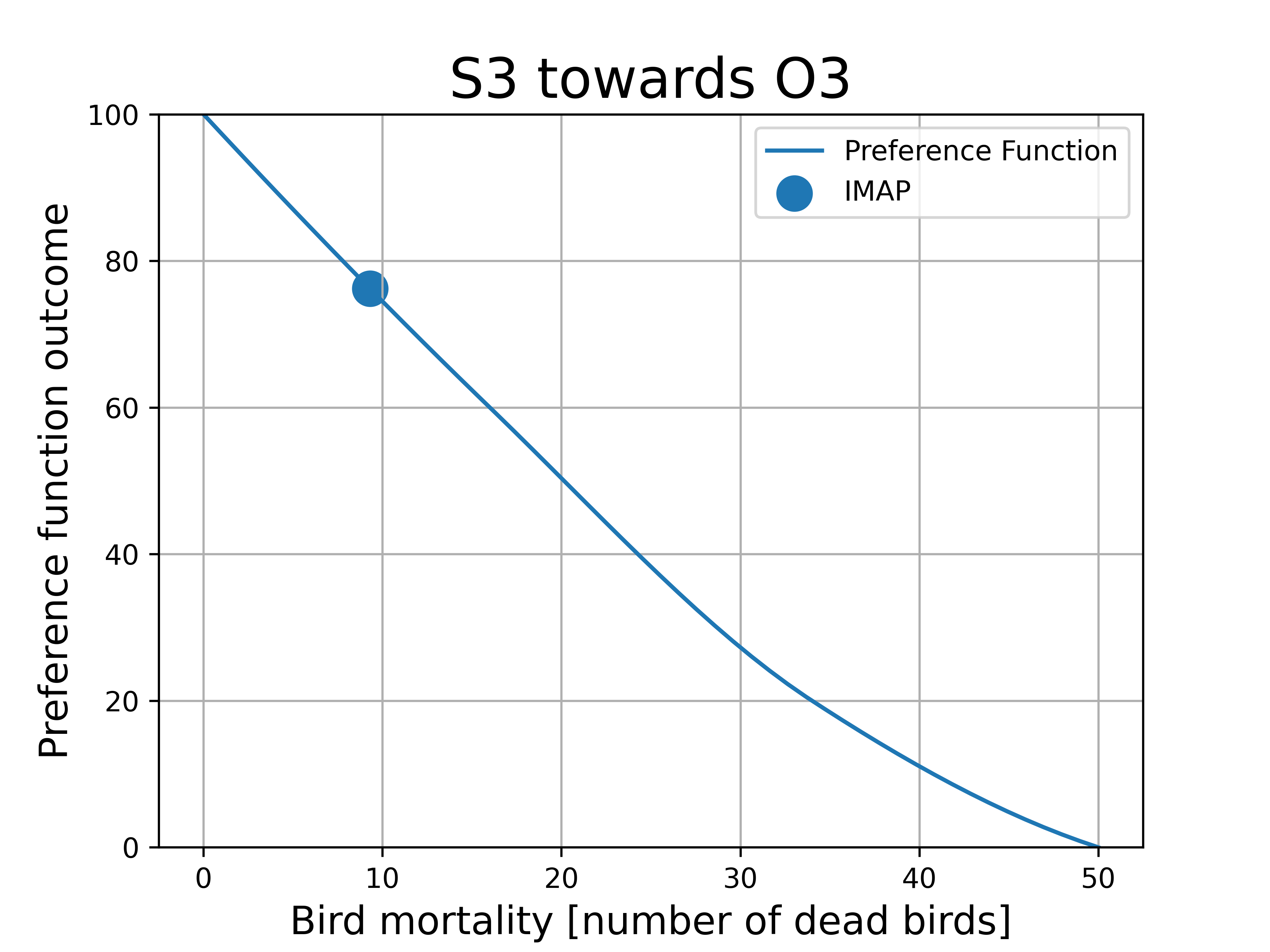}
    \end{minipage}
    \hfill
    \begin{minipage}[b]{0.40\linewidth}
        \includegraphics[width=\linewidth]{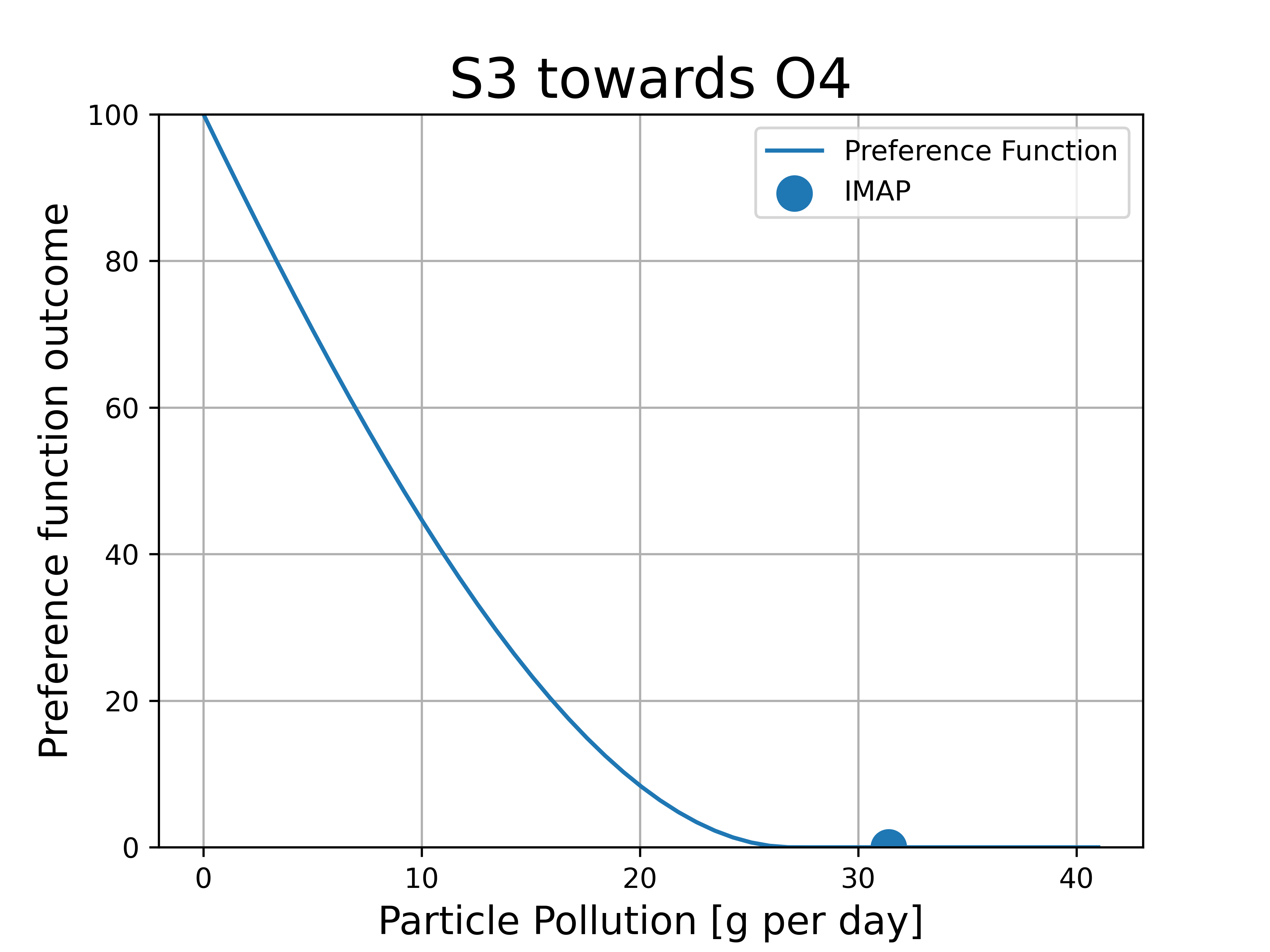}
    \end{minipage}
    \caption{Preference curves for $S_3$ towards $O{1..4}$.}
    \label{fig:4x4_S3}
\end{figure}

\begin{figure}[H]
    \centering
    \begin{minipage}[b]{0.40\linewidth}
        \includegraphics[width=\linewidth]{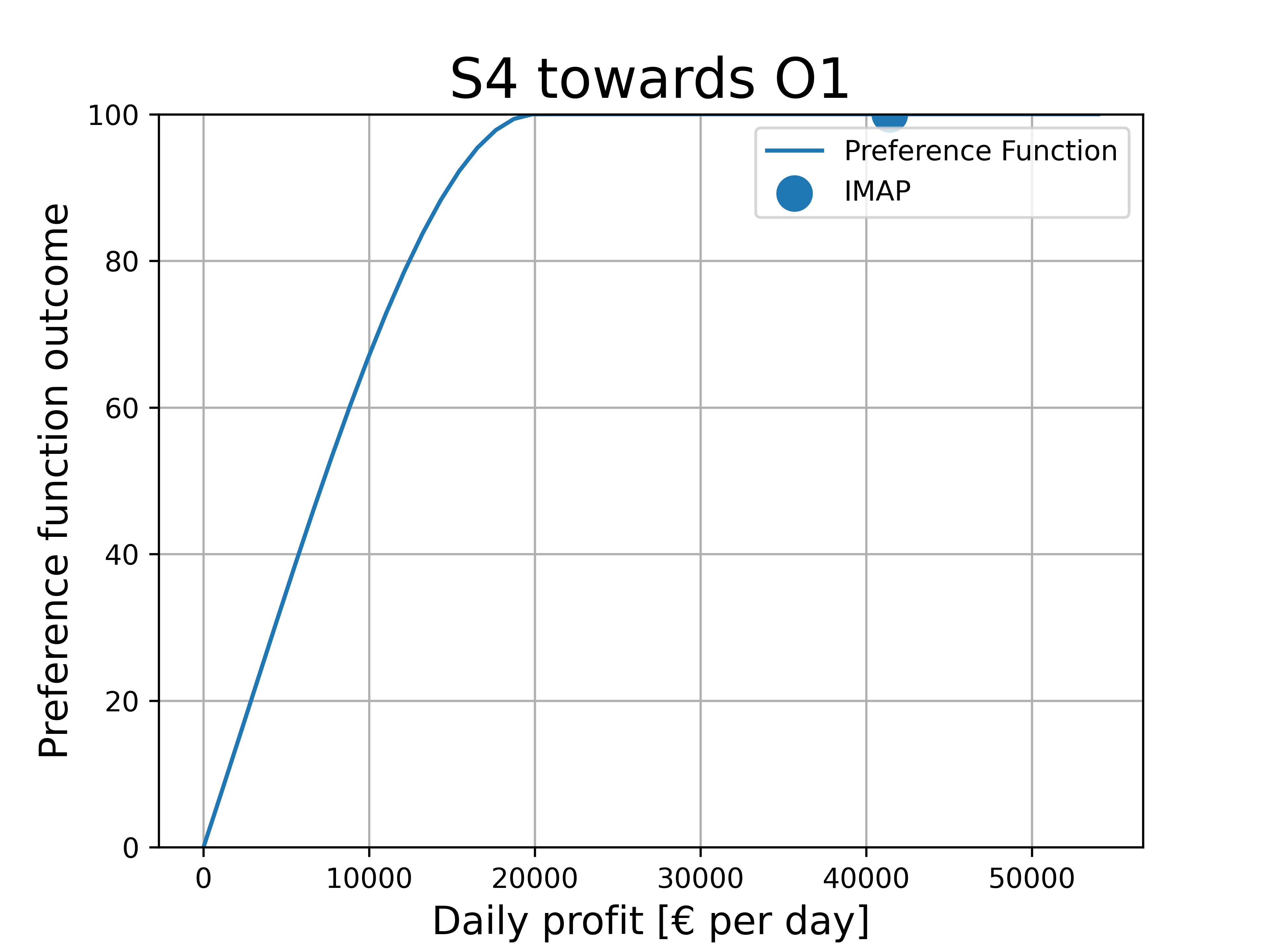}
    \end{minipage}
    \hfill
    \begin{minipage}[b]{0.40\linewidth}
        \includegraphics[width=\linewidth]{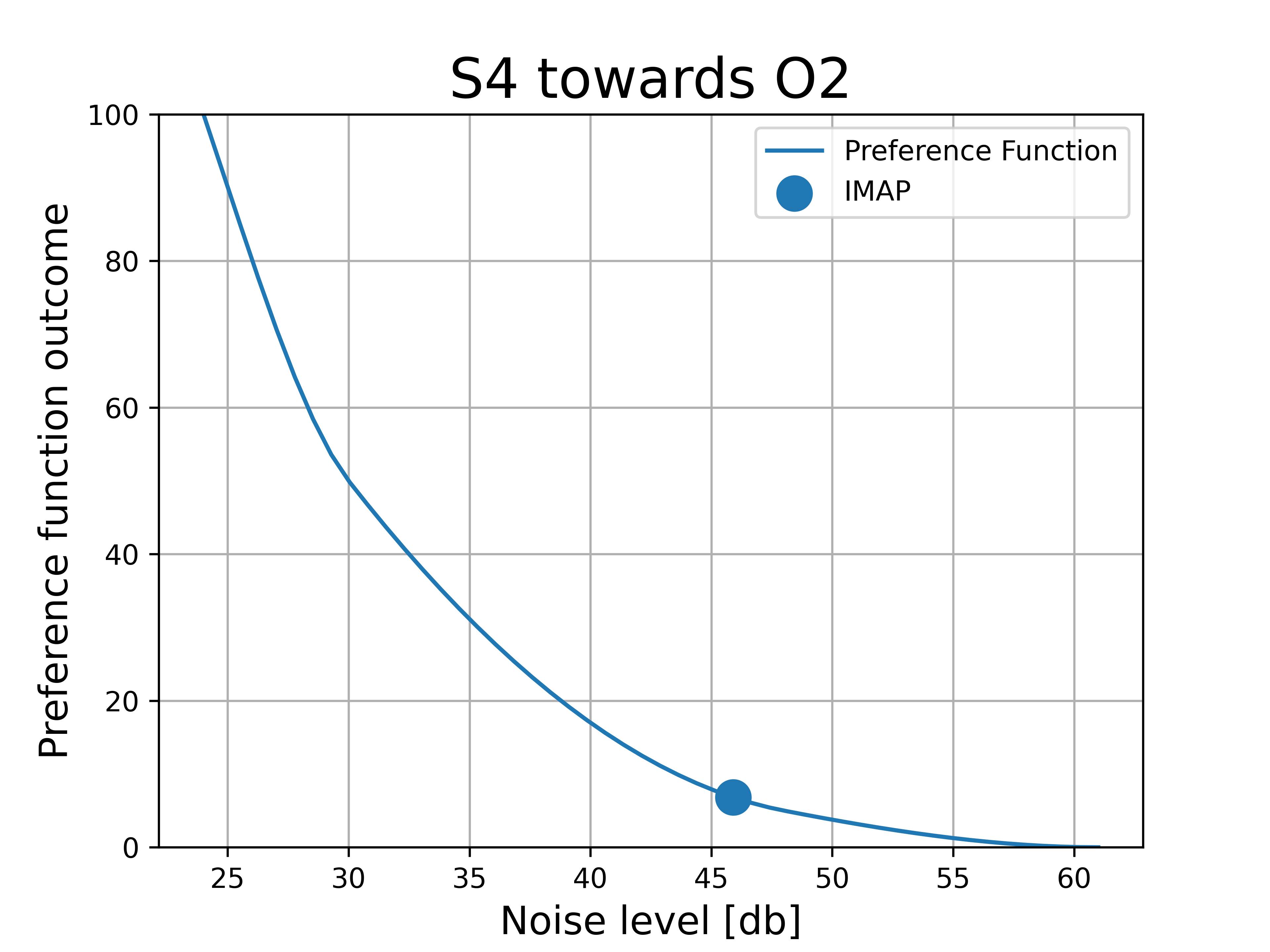}
    \end{minipage}
    
    \vspace{0.5cm} 

    \begin{minipage}[b]{0.40\linewidth}
        \includegraphics[width=\linewidth]{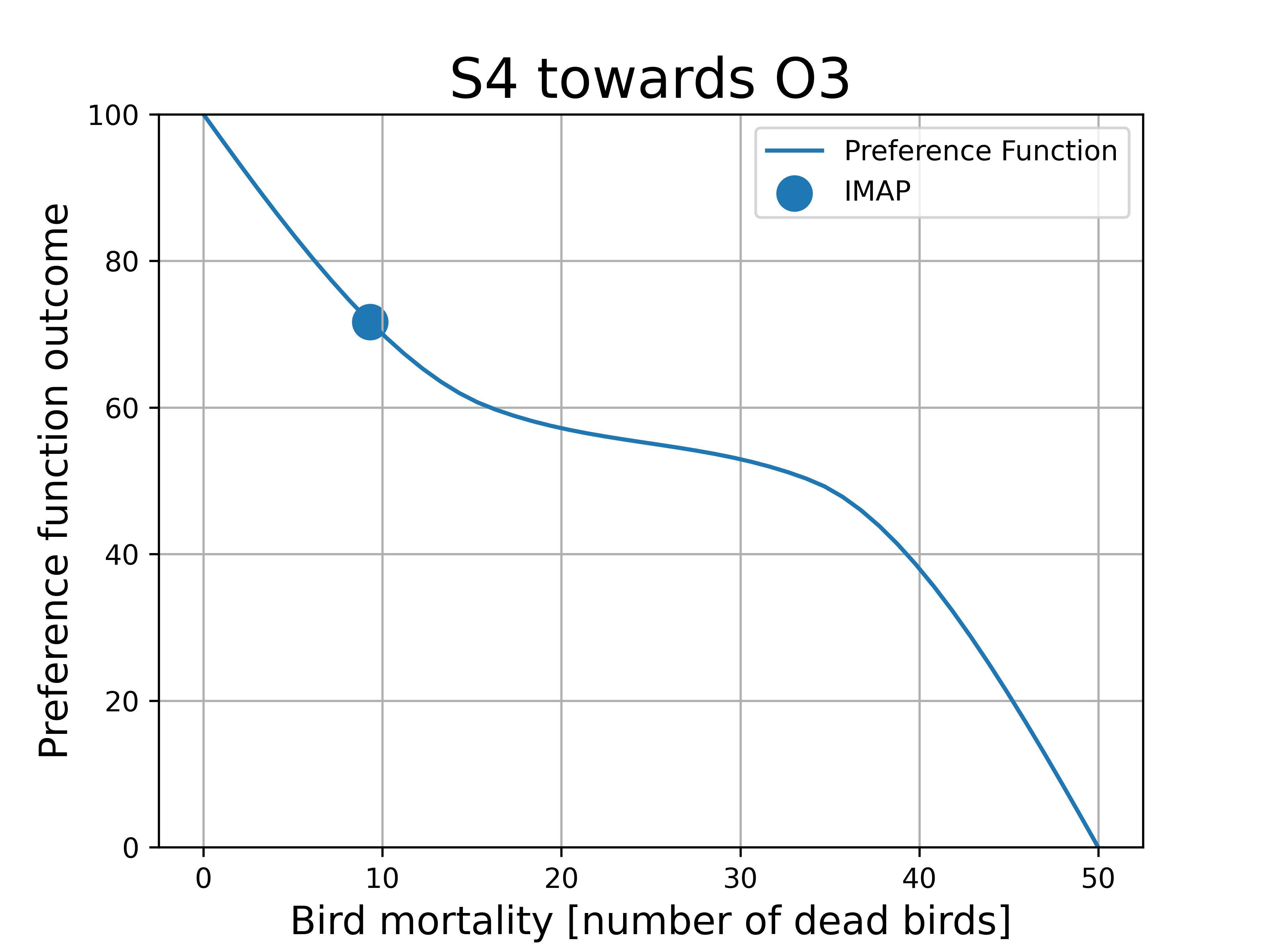}
    \end{minipage}
    \hfill
    \begin{minipage}[b]{0.40\linewidth}
        \includegraphics[width=\linewidth]{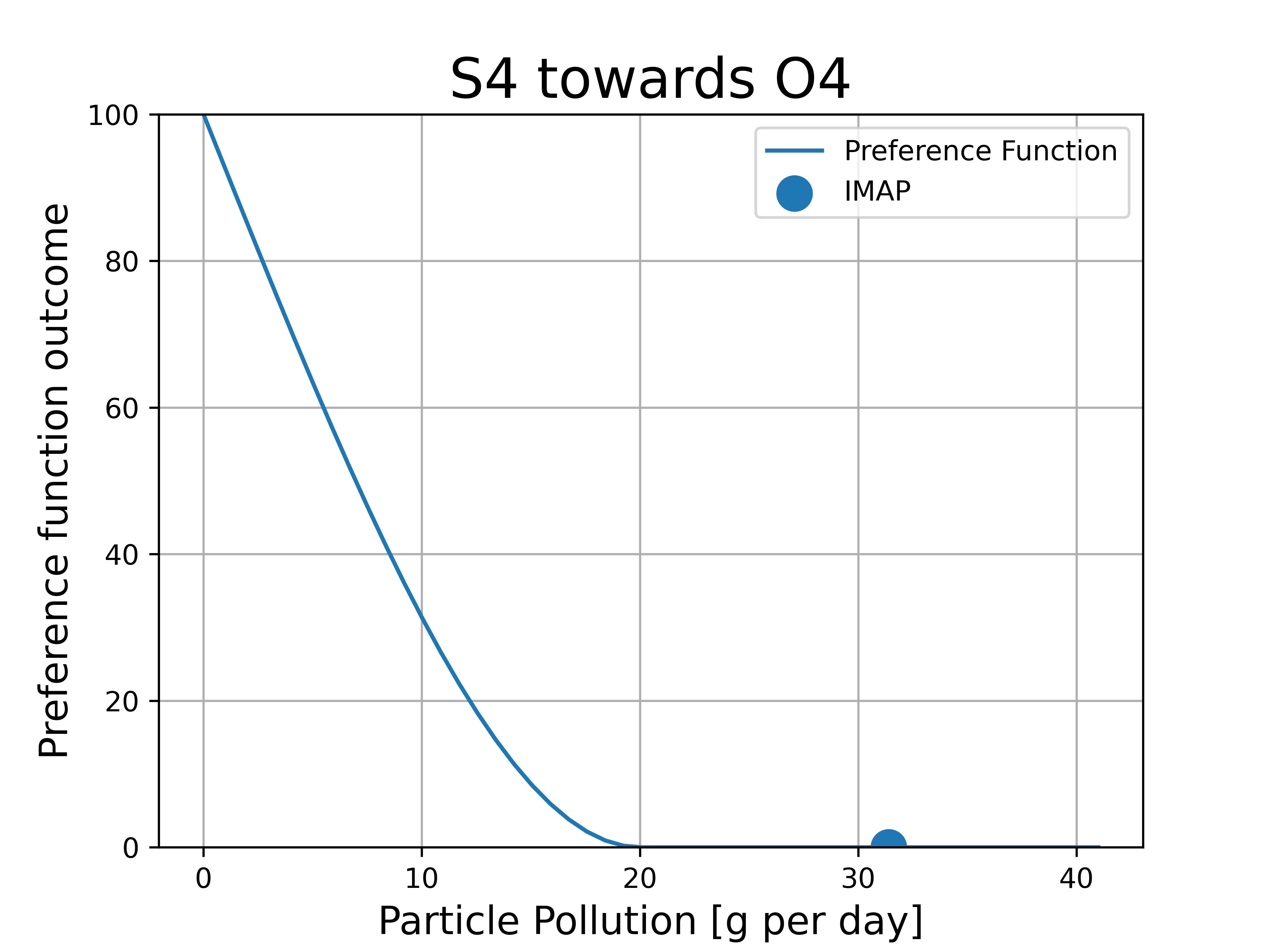}
    \end{minipage}
    \caption{Preference curves for $S_4$ towards $O{1..4}$.}
    \label{fig:4x4_S4}
\end{figure}

\section{Conclusion \& further development}
The working and added value of the complex systems design methodology Odesys have been demonstrated in an actual Dutch stalemate windfarm development problem. The dynamic tool, called Preferendus, has been used to confront socio-technical conflicts and move toward a collaborative 'yes.' Depending on the specific technical parameter settings and the combined interests of stakeholders, this has been partly achievable. For the Oss wind farm development, it was found that the 4x4 problem could be solved using MODO/IMAP, based on the current conjoint stakeholder judgment, for both linear and non-linear cases. However, in several cases, the solution proved infeasible due to physical limitations, as the required wind speeds are simply impossible. Moreover, when there is no dominant municipality, yet an equal partnership, or when there is no coalition for the energy objective, the solution is also non-viable, as the achievable energy benefits fall significantly below the necessary levels. These findings indicate that achieving a solution that is both realistic and desirable requires expanding the problem’s system boundaries to a zoomed-out solution space. This zoomed-out systems approach allows the energy transition challenge to be addressed through multiple viable alternatives rather than a single prescribed solution. This multi-energy problem can also be addressed using the Odesys methodology by allowing one of the design variables to represent different options (out of scope for this paper).

In addition to the previous specific Oss windfarm conclusions, the following generic conclusions can be drawn: 
\begin{itemize}
   \item  Following Odesys' associative and integrative design principle of diaduction, it is shown how to translate a stalemate problem into an IMAP framework. While this problem could not be fully dissolved, it was resolvable by using the IMAP based Preferendus to transform socio-technical conflicts into a co-created ‘yes,’ effectively conjoining stakeholder interests with practical realities to achieve the highest aggregated preferred value. In this way Odesys and its design engine have proven capable of revolutionizing decision-making, shifting it towards a truly direct-democratic form.
   \item It is shown that wishful thinking took precedence over realism in untangling this stalemate. However, physical feasibility ultimately brought the situation back to reality, echoing the well-known quote by natural scientist Richard Feynman: ‘For a successful technology, reality must take precedence over stakeholder relations, for nature cannot be fooled’.
   \item It is shown that when an individual stakeholder is willing to set aside purely self-interested motives, it becomes possible to transform the conflict into a best-fit result for all stakeholders involved. This aligns with the social law articulated by the spiritual scientist Rudolf Steiner: 'The well-being of a group of people working together is greater the less the individual claims the work outcome solely for their own interest.' Although in this design application no true best-fit solution for common purpose could be reached —since the outcomes were best-fit but not feasible or viable—, it can be concluded that the Odesys approach led to the principle: 'Better to turn back halfway than be completely lost.'
\end{itemize}
In addition to the specific and general conclusions mentioned above, the following recommendations are provided for further development:
\begin{itemize}
    \item The decision-making model needs to be expanded to include several energy alternatives in addition to just the wind energy variant. Only then can a desirable, feasible and technically feasible energy transition scenario be found.
    \item The technical parameters and design performance functions that describe the behaviour of this wind farm need to be further tested in practice to make the Preferendus results more reliable and usable in a real stalemate negotiations (i.e., improve structured expert judgment). 
   \item The social input, the preference functions describing the stakeholders interests and the steps in the social cycle should be studied in more detail (i.e., improve structured stakeholder judgment). It is also recommended to synchronize these with Friedrich Glasl's model of conflict escalation and with Roger Fisher's Getting to Yes principles (see \citeshortt{Wolfert2023}).
   \item A specific Preferendus stalemate solver module could be developed that can be used to determine where in the design problem the least amount of concession should be made to achieve a desired and achievable result. This could include the addition of negotiable constraints or preference elicitation.
   \item The process of going through an integrated socio-technical cycle needs to be further developed, tested and validated in real-life practice. It is recommended to using the latest insights from Otto Scharmer's U-model theory and practice. In particular, the concepts of 'dialoguing in the now' and 'idealised design' deserve specific process improvements for the Odesys U-model and its open-loop learning approach: i.e., a DMIII approach moving a step beyond the single MI and double MII loops learning from Chris Argyris and Donald Sch\"on.  
\end{itemize}


\section*{Acknowledgements}
The authors would like to thank the following Open Design Systems developers: Anne-Karien van der Stee, Marieke Schrievers, and Boudewijn Mol, for advancing the work on conjoint preference elicitation and structured stakeholder judgment. We also extend our thanks to Boskalis and Microsoft for facilitating these developments. Finally, the authors wish to thank Dr. Ruud Binnekamp of Delft University of Technology for his valuable reflections and guidance.

\section*{Disclosure statement}
The authors report that there are no competing interests to declare.

\section*{Data availability statement}\label{data availability}

Separate Deepnote links are available for each of the three case parts. This allows the reader to dynamically ‘play' with the model so that it starts 'talking back'. These three app-links are:

\begin{enumerate}
    \item 2x2 single-interest \& linear: 
\url{https://deepnote.com/app/odesys-toolbox-2/Windfarm-basic-6a6010c5-ebd0-4c0f-8892-ec833cfead85}

     \item 4x4 single-interest \& linear: 
\url{https://deepnote.com/app/odesys-toolbox-2/Windfarm-4x4-linear-d60d004f-68fe-4046-9e69-2887324d6760}
    
    \item 4x4 multiple interest \& non-linear: 
\url{https://deepnote.com/app/odesys-toolbox-2/Windfarm-extended-13678eab-714e-487c-a68a-934298a161dc}
\end{enumerate}

\bibliographystyle{apacite}
\bibliography{MiCo}

\end{document}